\documentclass[12pt]{article}
\usepackage{epsfig}
\usepackage{axodraw}
\usepackage{a4}
\usepackage{latexsym}

\usepackage{pslatex}
\usepackage[latin1]{inputenc}
\usepackage[T1]{fontenc}

\newcommand{\beq}{\begin{equation}}
\newcommand{\eeq}{\end{equation}}
\newcommand{\bea}{\begin{eqnarray}}
\newcommand{\eea}{\end{eqnarray}}

\newcommand{\MSb}{$\overline{\mbox{MS}}$}
\newcommand{\as}{\alpha_{\rm s}}
\newcommand{\DD}{{\cal D}}

\begin{document}
\setlength{\parskip}{0.3cm}
\setlength{\baselineskip}{0.55cm}

\def\plus{{\!+\!}}
\def\minus{{\!-\!}}
\def\z#1{{\zeta_{#1}}}
\def\ca{{C_A}}
\def\cf{{C_F}}
\def\nf{{n^{}_{\! f}}}
\def\S(#1){{{S}_{#1}}}
\def\Ss(#1,#2){{{S}_{#1,#2}}}
\def\Sss(#1,#2,#3){{{S}_{#1,#2,#3}}}
\def\Ssss(#1,#2,#3,#4){{{S}_{#1,#2,#3,#4}}}
\def\Sssss(#1,#2,#3,#4,#5){{{S}_{#1,#2,#3,#4,#5}}}
\def\Npm{{{\bf N_{\pm}}}}
\def\Npmi{{{\bf N_{\pm i}}}}
\def\Nminus{{{\bf N_{-}}}}
\def\Nplus{{{\bf N_{+}}}}
\def\Nminustwo{{{\bf N_{-2}}}}
\def\Nplustwo{{{\bf N_{+2}}}}
\def\Nminusthree{{{\bf N_{-3}}}}
\def\Nplusthree{{{\bf N_{+3}}}}

\def\pqq(#1){p_{\rm{qq}}(#1)}
\def\Li(#1,#2){{\rm{Li}}_{#1}(#2)}
\def\H(#1,#2,#3,#4){{\rm{H}}_{#1,#2,#3}(#4)}

\begin{titlepage}
\noindent
TTP$\,$02-22 \hfill {\tt hep-ph/0209100}\\
DESY 02-128 \\
NIKHEF 02-007 \\
September 2002 
\vspace{1.2cm}
\begin{center}
\Large
{\bf Non-Singlet Structure Functions at Three Loops:} \\
\vspace{0.15cm}
{\bf Fermionic Contributions} \\
\vspace{1.8cm}
\large
S. Moch$^{\, a,b}$, J.A.M. Vermaseren$^{\, c}$ and A. Vogt$^{\, c}$\\
\vspace{1.3cm}
\normalsize
{\it $^a$ Institut f{\"u}r Theoretische Teilchenphysik\\
\vspace{0.5mm}
Universit{\"a}t Karlsruhe, D--76128 Karlsruhe, Germany}\\
\vspace{0.5cm}
{\it $^b$Deutsches Elektronensynchrotron DESY \\
\vspace{0.5mm}
Platanenallee 6, D--15738 Zeuthen, Germany$\,$\footnote
{Address after September 1st, 2002}}\\
\vspace{0.5cm}
{\it $^c$NIKHEF Theory Group \\
\vspace{0.5mm}
Kruislaan 409, 1098 SJ Amsterdam, The Netherlands} \\
\vspace{3.2cm}
\large
{\bf Abstract}
\vspace{-0.2cm}
\end{center}
We compute the fermionic ($\nf$) contributions to the flavour 
non-singlet structure functions in unpolarized electromagnetic 
deep-inelastic scattering at third order of massless perturbative QCD.
Complete results are presented for the corresponding $\nf$-parts of the
three-loop anomalous dimension and the three-loop coefficient functions 
for the structure functions $F_2$ and $F_L$. Our results agree with all 
partial and approximate results available in the literature. The 
present calculation also facilitates a complete determination of the 
threshold-resummation parameters $B_2$ and $D_2^{\,\rm DIS}$ of which
only the sum was known so far, thus completing the information required
for the next-to-next-to-leading logarithmic resummation. We find that 
$D_2^{\,\rm DIS}$ vanishes in the \MSb\ scheme.
\vfill
\end{titlepage}
%
%
\section{Introduction}
\label{sec:introduction}
%
%
Structure functions in deep-inelastic scattering (DIS) form the 
backbone of our knowledge of the proton's parton densities, which are 
indispensable for analyses of hard scattering processes at proton--%
(anti-)proton colliders like the {\sc Tevatron} and the future LHC. 
Structure functions are also among the quantities best suited for 
precisely measuring the strong coupling constant $\as$. Over the past
twenty years DIS experiments have proceeded to a high (few-percent) 
accuracy and a wide kinematic coverage~\cite{Hagiwara:pw}. More results,
especially at high scales $Q^2$, can be expected from the forthcoming
high-luminosity phase of the electron--proton collider HERA at DESY.
On the theoretical side, at least the next-to-next-to-leading order 
(NNLO) corrections of perturbative QCD need to be taken into account in 
order to make full use of these measurements and to make quantitatively
reliable predictions for hard processes at hadron colliders.

The calculation of NNLO processes in perturbative QCD is far from easy. 
For deep-inelastic structure functions, in particular, the current 
situation is that, while the coefficient functions are known to two 
loops~\cite{vanNeerven:1991nn,Zijlstra:1991qc,Zijlstra:1992kj,%
Zijlstra:1992qd,Moch:1999eb}, only six/seven integer Mellin moments of 
the corresponding three-loop anomalous dimensions have been computed 
for lepton--hadron \cite{Larin:1994vu,Larin:1997wd,Retey:2000nq} and 
lepton--photon DIS \cite{Moch:2001im}, together with the same moments 
of the three-loop coefficient functions. The hadronic results have been 
employed, directly~\cite{Kataev:1997nc,Kataev:1999bp,Santiago:1999pr,%
Santiago:2001mh} and indirectly~\cite{Alekhin:2001ih,Martin:2002dr} via 
$x$-space approximations constructed from them~\cite
{vanNeerven:1999ca,vanNeerven:2000uj,vanNeerven:2000wp}, to improve the 
data analysis and some hadron-collider predictions. However, the number 
of available moments is rather limited, and hence these results cannot 
provide sufficient information at small values of the Bjorken variable 
$x$.

For the complete information one needs to obtain either all even or odd 
(depending on the quantity under consideration) Mellin moments, or do 
the complete calculation in Bjorken-$x$ space. We have adopted the 
first approach. Following the formalism of ref.~\cite{Larin:1991zw,%
Larin:1991tj,Larin:1994vu,Larin:1997wd} to obtain the lower fixed 
moments, we have used recursive methods to extend the calculation to 
all values of the Mellin moment $N$. This is by no means trivial, 
since before the start of the calculation the mathematics of the answer 
was still poorly understood~\cite{Vermaseren:2000we}. 
Hence it was first necessary to develop the understanding of harmonic 
sums~\cite{Gonzalez-Arroyo:1979df,Gonzalez-Arroyo:1980he,%
Vermaseren:1998uu,Blumlein:1998if} and harmonic polylogarithms~\cite
{Goncharov,Borwein,Remiddi:1999ew}. In addition the Mellin transforms 
and the inverse Mellin transforms from Bjorken-$x$-space to Mellin 
space and back had to be solved~\cite{Remiddi:1999ew}. 
These conceptual problems have been overcome and the method has been 
shown to work in a complete re-calculation of the two-loop coefficient 
functions~\cite{Moch:1999eb}.

The concept of working in Mellin space is not new. This method was 
already used in the early QCD papers~\cite{Gross:1973rr,Georgi:sr,%
Gross:cs}. But even in the case of the two-loop anomalous 
dimensions it was still possible to do the resulting sums in a rather 
direct manner~\cite{Floratos:1977au,Gonzalez-Arroyo:1979df}. This 
changed with the two-loop evaluation of $\sigma_L/\sigma_T$ in which 
Kazakov and Kotikov \cite{Kazakov:1988jk,Kazakov:1990jm} managed to 
obtain some of the integrals via recursion relations or first order 
difference equations.

In this article, we present the fermionic ($\,\nf\,$) corrections to 
the flavour non-singlet structure functions in electromagnetic DIS at 
the three-loop level. This includes the three-loop anomalous dimensions 
which are needed for the completion of the NNLO calculation, as well as 
the three-loop coefficient functions which are, at least at large $x$, 
the most important contribution to the next-to-next-to-next-to-leading 
order (N$^{3}$LO) correction \cite{vanNeerven:2001pe}. Of course, the 
$\nf$-part is not the complete calculation. Yet we decided to present 
it already now, because the complete calculation (including the singlet 
part) will still take quite some (computer)$\,$time. Thus, for the 
time being, NNLO calculations of the Drell-Yan process~\cite
{Hamberg:1990np,Harlander:2002wh} (which are relevant for luminosity
monitoring at {\sc Tevatron} and LHC~\cite{Wprod1,Wprod2,Wprod3}) and
of Higgs production~\cite{Harlander:2002wh,Anastasiou:2002yz} have to 
rely on parton distributions evolved with the approximate splitting 
functions of ref.~\cite{vanNeerven:2000wp}. Our present calculation 
provides a first check of the reliability of these approximations. 
More importantly, it turns out that already this calculation is 
sufficient to provide the last relevant missing information for the 
extension of the soft-gluon (threshold) resummation for 
DIS~\cite{Sterman:1987aj,Catani:1989ne,Catani:1991rp} to the 
next-to-next-to-leading logarithmic accuracy~\cite{Vogt:2000ci}. 
In fact, our result for the resummation parameter $D_2^{\,\rm DIS}$ is 
most intriguing and calls for further studies.

This article is organized as follows. In section~\ref{sec:method} we 
outline those parts of the calculation, which differ from previous
two-loop~\cite{Moch:1999eb} and fixed-$N$ three-loop~\cite
{Larin:1994vu,Larin:1997wd,Retey:2000nq} analyses. The present 
calculation does not yet involve the full complexity of the method, as
the most difficult diagram topologies do not occur. Therefore we 
postpone a full account to a later publication. In section~\ref
{sec:results} we present our explicit even-$N$ Mellin-space results,
except for the rather lengthy expressions for the three-loop coefficient
functions which are deferred to appendix \ref{sec:appendix}. The 
corresponding three-loop quantities in $x$-space can be found in section
\ref{sec:xresults}. Here, instead of writing down the cumbersome exact
expressions for the coefficient functions, we follow the procedure 
applied in refs.~\cite{vanNeerven:1999ca,vanNeerven:2000uj} to the two-%
loop coefficient functions, and provide approximate parametrizations 
which are compact and sufficiently accurate for all numerical 
applications. In section~\ref{sec:sresults} we then discuss the 
implications of our calculation on the soft-gluon resummation, before 
we summarize our results in section~\ref{sec:summary}.
%
%
\section{Method} 
\label{sec:method}
%
%
Because we are considering only the non-singlet structure functions in this 
paper, the method for the calculation of their moments closely 
follows ref.~\cite{Larin:1994vu}. Hence there is not much need to explain the 
physics of the method here again. Thus we will discuss only the differences 
introduced by the fact that we now compute all moments simultaneously as a 
function of the moment number $N$. 
Since $N$ is not a fixed constant, we cannot resort to the techniques of 
ref.~\cite{Larin:1994vu}, where the Mincer program~\cite{Gorishnii:1989gt,%
Larin:1991fz} was used as the tool to solve the integrals. Instead, we will 
have to introduce new techniques. However, we can give $N$ a positive integer 
value at any point of the derivations and calculations, after which the Mincer 
program can be invoked to verify that the results are correct. From a practical 
point of view this is the most powerful feature of the Mellin-space approach, 
as it greatly simplifies the checking of all programs.

Similar to the fixed-$N$ computations of refs.~\cite
{Larin:1997wd,Retey:2000nq}, the diagrams are generated automatically with a 
special version of the diagram generator QGRAF~\cite{Nogueira:1991ex}.
For all the symbolic manipulations of the formulae we use the latest version of 
the program FORM~\cite{Vermaseren:2000nd}.
The calculation is performed in dimensional regularization~\cite
{'tHooft:1972fi,Bollini:1972ui,Ashmore:1972uj,Cicuta:1972jf} with 
$D=4-2\epsilon$. Hence the unrenormalized Mellin-space results will be 
functions of $\epsilon,\: N,$ and the values $\zeta_{3,\ldots, 5}$ of the 
Riemann $\zeta$-function. The renormalization is carried out in the \MSb-scheme
\cite{'tHooft:1973mm,Bardeen:1978yd} as described in ref.~\cite{Larin:1994vu}.

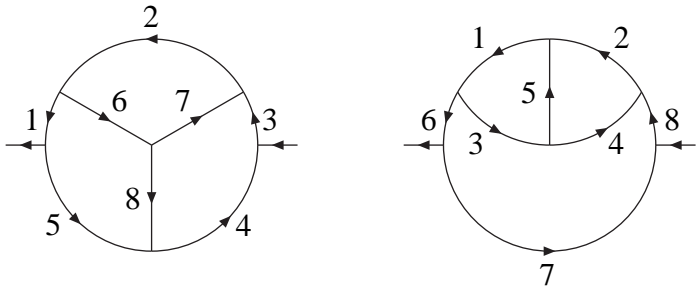
\begin{figure}[htb]
\begin{center} 
\begin{picture}(260,110)(0,0)
\ArrowLine(5.4,65)(40,45) 
\ArrowLine(40,45)(74.6,65)
\ArrowLine(40,45)(40,5)
\ArrowArc(40,45)(40,270,360)
\ArrowArc(40,45)(40,180,270)
\ArrowArc(40,45)(40,0,30)
\ArrowArc(40,45)(40,30,150)
\ArrowArc(40,45)(40,150,180)
\ArrowLine(0,45)(-15,45) 
\ArrowLine(95,45)(80,45)
\Text(-2,55)[r]{1}
\Text(40,90)[b]{2}
\Text(82,55)[l]{3}
\Text(72,15)[l]{4}
\Text(6,15)[r]{5}
\Text(25,62)[l]{6}
\Text(55,62)[r]{7}
\Text(36,25)[r]{8}
\ArrowLine(190,45)(190,85)
\ArrowArc(190,45)(40,180,360)
\ArrowArc(190,45)(40,0,30)
\ArrowArc(190,45)(40,30,90)
\ArrowArc(190,45)(40,90,150)
\ArrowArc(190,45)(40,150,180)
\ArrowArc(190,85)(40,210,270)
\ArrowArc(190,85)(40,270,330)
\ArrowLine(150,45)(135,45) 
\ArrowLine(245,45)(230,45)
\Text(161,82)[lb]{1}
\Text(221,82)[rb]{2}
\Text(160,50)[lt]{3}
\Text(219,50)[rt]{4}
\Text(186,65)[r]{5}
\Text(148,55)[r]{6}
\Text(190,0)[t]{7}
\Text(234,55)[l]{8}
\end{picture}
\end{center}
\vspace{-3mm}
\caption{\label{pic:topo} The topologies $\rm BE$ (left) and O4 (right) of 
propagator-type diagrams, with the line numbering as employed in figs.~\ref
{pic:benz} and \ref{pic:o4} below. The external lines carry the momentum $Q$.}
\end{figure}

We distinguish three categories of diagrams: complete diagrams, composite 
building blocks and basic building blocks. A complete diagram is a Feynman 
diagram with all its structure like traces and dotproducts in the numerator. 
Such a diagram may lead to a large number of more fundamental integrals that 
cannot be reduced by considerations of momentum conservation only. For the 
understanding of composite and basic building blocks, one has to realize that 
in the framework of the operator-product expansion we eventually have to take 
$N$ derivatives with respect to the parton momentum $P$ after which $P$ is put
equal to zero. This projects out the $N$-th Mellin moment~\cite
{Gorishnii:1983su} and it effectively amputates the legs of the parton, leaving
us with propagator-type diagrams. 
Therefore, we define the topology of a diagram as the propagator topology when 
the $P$-momentum legs have been amputated. The three-loop propagator topologies 
of the BE (Benz) type  and of the O4 type are shown in fig.~\ref{pic:topo}.
For the notation we refer to refs.~\cite{Gorishnii:1989gt,Larin:1991fz}. The 
external lines in the propagator topology are referred to as $Q$.

When the numbers of the position of the incoming and outgoing momenta have
been attached we are referring to subtopologies. For instance, BE$_{13}$ is a 
subtopology of type BE in which the momentum $P$ comes in in line 1 and leaves 
in line 3, assuming the numbering of the BE topology as in fig.~\ref{pic:topo}. 
We define basic building blocks (BBB) as integrals in which both the incoming 
and the outgoing $P$-momentum are attached to the same line as, e.g., in 
BE$_{22}$. In composite building blocks (CBB), on the other hand, the incoming 
and the outgoing $P$-momentum are attached to different lines, as in the
case of BE$_{13}$ mentioned above. 
Of course we could have introduced names for all these three-loop four-point 
functions, but since eventually the $P$-momentum legs get amputated the above 
notation seems the clearest scheme. In this way, it is only a small step to an 
easy pictorial representation of the integrals as used in ref.~\cite
{Moch:1999eb}.

\begin{figure}[thb]
\begin{center} \begin{picture}(260,85)(0,-5)
\SetColor{Red}  
\Gluon(35,0)(0,0){4}{4} 
\Gluon(70,60)(70,30){4}{4} 
\ArrowLine(-15,-10)(0,0)
\ArrowLine(35,0)(70,30)
\ArrowLine(100,60)(115,-10)
\ArrowLine(70,60)(100,60)
\SetColor{Black} 
\Photon(100,60)(115,70){2}{2}
\Photon(0,60)(-15,70){2}{2}
\Gluon(0,30)(35,30){4}{4}
\ArrowLine(0,60)(70,60)
\ArrowLine(70,30)(35,30)
\ArrowLine(35,30)(35,0)
\ArrowLine(0,0)(0,30)
\ArrowLine(0,30)(0,60)
\Text(-5,45)[r]{6}
\Text(-5,15)[r]{1}
\Text(17,-8)[t]{1}
\Text(30,15)[r]{5}
\Text(17,38)[b]{4}
\Text(52,36)[b]{3}
\Text(56,11)[lt]{2}
\Text(78,45)[l]{8}
\Text(35,66)[b]{7}
\SetColor{Red}  
\Gluon(185,0)(150,0){4}{4} 
\Gluon(250,20)(220,20){4}{4} 
\ArrowLine(135,-10)(150,0)
\ArrowLine(185,0)(220,20)
\ArrowLine(250,60)(250,20)
\ArrowLine(250,20)(265,-10)
\SetColor{Black} 
\Photon(250,60)(265,70){2}{2}
\Photon(150,60)(135,70){2}{2}
\Gluon(150,30)(185,30){4}{4}
\ArrowLine(150,60)(250,60)
\ArrowLine(220,20)(185,30)
\ArrowLine(185,30)(185,0)
\ArrowLine(150,0)(150,30)
\ArrowLine(150,30)(150,60)
\Text(145,45)[r]{6}
\Text(145,15)[r]{1}
\Text(167,-8)[t]{1}
\Text(180,15)[r]{5}
\Text(167,38)[b]{4}
\Text(204,6)[lt]{2}
\Text(204,29)[lb]{3}
\Text(200,66)[b]{7}
\Text(235,12)[t]{8}
\Text(256,40)[l]{8}
\end{picture}
\end{center}
\caption{\label{pic:o4} The diagrams of topology O4, which contribute to the 
 fermionic part of the non-singlet structure functions $F_2$ and $F_L$. These
 diagrams are examples of the subtopologies O4$_{\rm 1Q}$ and O4$_{18}$.}
\end{figure}
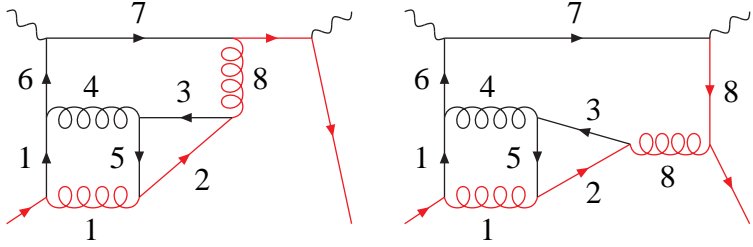

For the calculation of the $n_f$-parts of the non-singlet structure functions
$F_2$ and $F_L$, one does not need to consider all three-loop topologies. In
fact, the only genuine three-loop subtopologies one has to solve are of the 
BE-type and of the O4-type, with two complete diagrams of either type entering
the calculation. These diagrams are displayed in figs.~\ref{pic:benz} 
and~\ref{pic:o4}.
 
\begin{figure}[thb]
\begin{center} \begin{picture}(260,85)(0,-5)
\SetColor{Red}  
\Gluon(35,0)(0,0){4}{4} 
\Gluon(70,60)(70,30){4}{4} 
\ArrowLine(-15,-10)(0,0)
\ArrowLine(35,0)(70,30)
\ArrowLine(100,60)(115,-10)
\ArrowLine(70,60)(100,60)
\SetColor{Black} 
\Photon(100,60)(115,70){2}{2}
\Photon(0,60)(-15,70){2}{2}
\Gluon(35,30)(35,60){4}{4}
\ArrowLine(0,0)(0,60)
\ArrowLine(0,60)(35,60)
\ArrowLine(35,60)(70,60)
\ArrowLine(70,30)(35,30)
\ArrowLine(35,30)(35,0)
\Text(-5,30)[r]{1}
\Text(17,-10)[t]{1}
\Text(54,11)[lt]{2}
\Text(78,45)[l]{3}
\Text(52,65)[b]{4}
\Text(17,65)[b]{5}
\Text(29,15)[r]{6}
\Text(52,36)[b]{7}
\Text(28,45)[r]{8}
\SetColor{Red}  
\Gluon(185,0)(150,0){4}{4} 
\Gluon(250,0)(220,0){4}{4}
\ArrowLine(135,-10)(150,0)
\ArrowLine(250,0)(265,-10)
\ArrowLine(185,0)(220,0)
\SetColor{Black} 
\Photon(250,60)(265,70){2}{2}
\Photon(150,60)(135,70){2}{2}
\Gluon(200,60)(200,30){4}{4}
\ArrowLine(150,0)(150,60)
\ArrowLine(150,60)(200,60)
\ArrowLine(200,60)(250,60)
\ArrowLine(250,60)(250,0)
\ArrowLine(220,0)(200,30)
\ArrowLine(200,30)(185,0)
\Text(145,30)[r]{1}
\Text(167,-10)[t]{1}
\Text(202,-6)[t]{2}
\Text(235,-10)[t]{3}
\Text(256,30)[l]{3}
\Text(225,65)[b]{4}
\Text(175,65)[b]{5}
\Text(189,17)[br]{6}
\Text(214,17)[lb]{7}
\Text(194,45)[r]{8}
\end{picture}
\end{center}
\caption{\label{pic:benz} The diagrams of topology $\rm BE$ which contribute to 
 the fermionic part of the non-singlet structure functions $F_2$ and $F_L$. The 
 subtopologies are ${\rm BE}_{1\rm Q}$ (left) and ${\rm BE}_{13}$ (right).}
\end{figure}
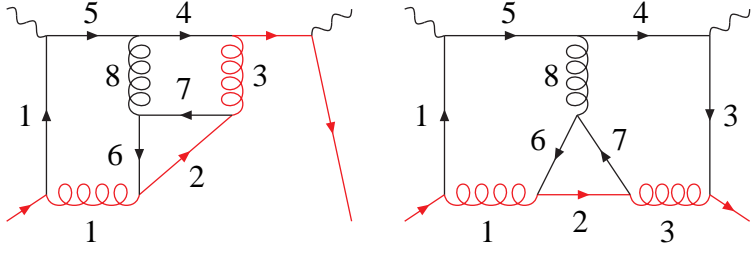
 
The main problem we are faced with as compared to the corresponding two-loop
calculation~\cite{Moch:1999eb} is that the necessary reduction equations become 
much more complicated. In addition the set of equations that was available at 
that moment was not maximal, and the structure of the more complicated 
topologies needs the maximal set of reduction equations. This involves some 
equations in which the vanishing of $P\cdot P$ is an issue that should be 
considered with care, a point that will be explained in full detail in a later
publication in which we have to deal with all topologies.

For this calculation we first studied the basic building blocks. Here it is 
easy to expand the single propagator that contains the momentum $P$ to 
sufficient powers in $P$ for the $N$-th moment. This number of powers can
be less than $N$ as there may be some powers of $P$ in the numerator 
already. It should be noted that if we have a power of $P\cdot Q$ in the 
numerator we are effectively computing the moment $N\!-\!1$ of the integral. At 
this point we write down all equations based on integration by parts, all 
scaling relations and all form-factor equations that can be constructed. 
Next follows a potentially difficult process in which we have to combine 
these relations to construct equations that can systematically bring the 
powers of the denominators in the integral down, either reducing them to 
zero or leaving them at a fixed unique value. When a line is eliminated 
a simpler topology is reached and we can refer to the reduction equations for 
that topology. This will eventually lead to a topology that is simple enough 
so that we can calculate the integral. 

A problem arises when a reduction equation cannot take the power $n$ of a 
denominator beyond one, because it is of the type $ n I(n+1) \rightarrow I(n)$. 
Also when the power of a propagator is not an integer, one can only try to 
reduce this power to a fixed value for which the line does not vanish. In 
these cases we leave the power either at one or at this fixed value 
(usually $1+\epsilon$) and we continue with the next propagator. Eventually 
we will have integrals in which only the line with the power that involves 
$N$ will not have a standard value. The $N$ will occur both in the power of 
the denominator $p\cdot p$ and in the numerator $2P\cdot p$ when $p$ is the 
momentum associated to this line. By this time our remaining equations 
which have also been treated to bring the powers of their progagators to 
standard values may 
have become rather lengthy. But we still need them to bring the difference 
of the powers in the numerator and the denominator to a fixed value, which 
can be either 0, 1 or 2, something we have to leave open for reasons explained 
below. Now there are several possibilities: If there is only one integral left 
of the type to be solved, the equation directly determines the solution. This 
is however rarely the case. Usually there are several terms left, each with a 
different power of $2 P\cdot Q$, leaving us with an equation of the type
\begin{equation}
\label{eq:deq}
a_0(N)\, I(N) + a_1(N)\,  I(N-1) + \ldots + a_m(N)\,  I(N-m) \: = \: G(N)
\end{equation}
in which the function $G$ refers to a potentially horrendous combination of 
integrals of simpler topologies. Eq.~(\ref{eq:deq}) defines a recursion 
relation or difference equation of order $m$. For the present calculation we 
did not have to go beyond order 2.
It should be noted that recently difference equations have been encountered by 
Laporta in refs.~\cite{Laporta:2000dc,Laporta:2001dd}.

These difference equations can be solved by making the ansatz that the solution 
will be a combination of harmonic sums. If the proper combination has been 
selected, the coefficients of the harmonic sums can be obtained by substituting 
the trial solution into the equation and solving the resulting, potentially 
large set of linear equations. There can be several thousand equations in such 
a system. Of course one has to have a solution for the function $G(N)$ in 
eq.~(\ref{eq:deq}) first, and $m\! -\! 1$ boundary values are required. Because 
these boundary values are basically fixed integer-$N$ moments, they can be 
obtained using the Mincer program. The need for knowing the function $G$ puts a 
rigid hierarchy in the order in which we have to treat the topologies.

Solving the difference equations is rather slow work. Hence we compute their 
solutions only once and tabulate the results. We usually do this for several 
values of the difference of the powers of the numerator and the denominator as 
mentioned above eq.~(\ref{eq:deq}). The reason behind this is that the 
equations we use for either raising or lowering this difference may contain a 
so-called spurious pole in $\epsilon$ when we try to bring this value to one. 
The concept of spurious poles is rather important. The rule of the triangle
\cite{Tkachov:1981wb,Chetyrkin:1981qh}, for instance,
can involve a factor proportional to 
$1/\epsilon$ when an integral is being reduced. Close inspection reveals that 
when powers of the loop momentum are present in the numerator, it is possible 
that more than one of such poles is generated before a denominator is removed. 
The resolved triangle~\cite{Tkachov:1984xk} shows, however, that it is possible 
to sum all contributions of such a reduction and that eventually there can be 
no more than one pole per eliminated line. This means that the extra poles 
should cancel between the many generated terms. But when we work only a to 
given cutoff in powers of $\epsilon$ (both for reasons of economy and 
because we cannot evaluate some integrals easily beyond certain powers in 
$\epsilon$) these temporary poles could spoil the final result in the same way 
as such things can happen in numerical calculations at fixed precision. We call 
them spurious poles, because in principle they can be avoided. One of the 
greatest difficulties in deriving reduction equations is to indeed avoid such 
spurious poles. For some integrals, no spurious-pole free formula could be 
found for the last reduction when bringing the difference of the power of the 
denominator and the numerator either from 2 to~1 or from 0 to~1. This problem
has been circumvented by solving the resulting difference equations for the 
three cases 0, 1 and 2 separately.

Since the evaluation of all the basic building block integrals that can occur 
requires very much computer time, we decided to tabulate all these integrals 
for the complicated topologies. This saves much computer time, because each 
integral is typically used many times. Also the more complicated topologies use 
many integrals of a less complicated type, hence their evaluation becomes much 
faster once the latter integrals have been tabulated.

The next step is the evaluation of the composite building blocks. Here again we 
first construct all possible equations. It turns out to be most economic to 
leave the propagators with the momentum $P$ unexpanded. The fact that 
eventually an expansion to $N$ powers of $P$ will take place then requires some 
special calculational rules. If for instance an equation is multiplied by 
$P\cdot Q$, we have to replace $N$ by $N\! -\! 1$ at the same time. Such rules 
can be major sources of errors. Hence it is very fortunate that at any moment 
we can decide that $N$ has a fixed value like four or five and then evaluate 
the integrals in the equation with the Mincer program to see whether it is 
still correct.

The equations are used to set up a reduction scheme that is similar to the one 
for the BBB's. If a line is eliminated we obtain either a simpler topology or a 
BBB. Otherwise we reduce the power of a denominator to a standard value and 
continue with the next line. In numerous cases a reduction can only be done by 
means of a difference equation~(\ref{eq:deq}). If this is a first order 
difference equation it can be solved directly, introducing one sum. Such sums 
are of a benign type and can be evaluated afterwards. If the difference 
equation is of a higher order we have to consider all more fundamental 
integrals first before we can solve the equation. The further reduction scheme 
becomes then rather complicated, but not impossible. On the average each 
subtopology can require several weeks of work before it has been completely 
solved by these methods.
 
Analogous to the BBB case discussed above, we have tabulated the more 
complicated CBB integrals entering the calculation. It is not only a matter of 
computer time that renders this necessary. Also the size of the intermediate 
expressions becomes a most relevant factor: if one is not careful, even a hard 
disk of 100 GBytes can become restrictive. In practice, already 
expressions significantly larger than 10 GBytes took too long for evaluation 
without further optimization. A careful hierarchy of tabulation managed to 
avoid these problems.

Having programs for all basic and composite building blocks renders the 
remainder of the calculation rather straightforward. The major difference to 
the fixed-moment calculations is now that we obtain much longer results due to 
the presence of the parameter $N$ in the answer. We have checked the 
correctness of each individual diagram for several values of $N$, by comparing 
with the results of a Mincer calculation. In addition we have compared the 
complete renormalized results with the results in the literature for the 
available values of $N$.
%
%
\section{Results in Mellin space}
\label{sec:results}
%
%
Here we present the $N$-space coefficient functions and the anomalous 
dimensions up to the third order in the renormalized coupling 
$\as$.  All results are given in the \MSb-scheme with the
renormalization scales identified with the physical hard scale $Q$.
Thus the perturbative expansion of the non-singlet coefficient functions
and anomalous dimensions can be written as
\bea
\label{eq:Cexp}
  C_{\,i,\rm ns}\left(\as,N\right) \! &= \!& \sum_{n=0}^\infty\,\: 
  \left(\frac{\as}{4\pi}\right)^{n}\: c^{\,(n)}_{\,i,\rm ns}(N) \:\: ,
\\ \label{eq:gexp}
 \gamma_{\,\rm{ns}}\left(\as,N\right) \, & = \!& \sum_{n=0}^\infty \, 
  \left(\frac{\as}{4\pi}\right)^{n+1} \gamma^{\,(n)}_{\,\rm ns}(N) 
\eea
with $i=2,L\,$ in eq.~(\ref{eq:Cexp}). There is no need to consider 
different choices of the scales, as they do not require functions 
beyond those introduced in eqs.~(\ref{eq:Cexp}) and (\ref{eq:gexp}). 
Recall that for $F^{}_2$ the $(n\! +\! 1)$-loop anomalous dimensions 
and the $n$-loop coefficient functions together form the 
N$^{\,\rm n}$LO approximation of (renormalization-group improved) 
perturbative QCD.

There are, actually, three different non-singlet combinations of 
coefficient functions and splitting functions. These combinations all
coincide at order ${\as}$, but they all differ beyond the second order.
Only the so-called `+'-combinations (involving sums over quarks and
antiquarks) are probed in electromagnetic DIS, hence only these 
quantities are addressed in the present article. Consequently our 
results below apply directly only to all even-integer values of $N$ 
from which, however, the results for arbitrary $N$ can be uniquely 
inferred by analytic continuation.

Our $N$-space results are expressed in terms of harmonic sums 
$S_{\vec{m}}(N)$. In the following all harmonic sums are understood to 
have the argument $N$, i.e., we employ the short-hand notation 
$S_{\vec{m}}\,\equiv\, S_{\vec{m}}(N)$. In addition we use operators 
$\Npm$ and $\Npmi$ which shift the argument $N$ of a given function by 
$\pm 1$ or a larger integer $i$,
\beq
\label{eq:shiftN}
  \Npm \, f(N) \: = \: f(N \pm 1) \:\: , \quad\quad
  \Npmi\, f(N) \: = \: f(N \pm i) \:\: .
\eeq

We normalize the trivial leading-order (LO) coefficient function
and recover, of course, the well-known result for the LO anomalous 
dimension~\cite{Gross:1973rr,Georgi:sr}
\bea
  c^{\,(0)}_{\,2,{\rm ns}}(N) & \! = \! & 
         1
\:\: ,\label{eq:c2qq0}
\\
  \gamma^{\,(0)}_{\,\rm ns}(N) & \! = \! & 
         \cf \* \big( 
            2 \* (\Nminus+\Nplus) \* \S(1) 
          - 3 
          \big)
\:\ .\label{eq:gqq0}
\eea
In our notation, the next-to-leading order (NLO) non-singlet coefficient
function for $F^{}_{2}$ \cite{Bardeen:1978yd} and the corresponding 
anomalous dimension \cite{Floratos:1977au,Gonzalez-Arroyo:1979df} read
\bea
  && \! c^{\,(1)}_{\,2,\rm ns}(N) \: = \: 
         \cf \* \big( 
            7\, \* \Nplus \* \S(1)
          + 2 \* \S(1)
          - 9
          + (\Nminus+\Nplus) \* [
            2 \* \Ss(1,1)
          - 3 \* \S(1)
          - 2 \* \S(2)
          ]  
          \big)
,\label{eq:c2qq1} 
\\[1mm]
  &&\gamma^{\,(1)}_{\,\rm ns}(N) \:\: = \:  
  4\, \* \ca \* \cf \* \bigg(
            2\, \* \Nplus \* \S(3)
          - {17 \over 24}
          - 2 \* \S(-3)
          - {28 \over 3} \* \S(1)
          + (\Nminus+\Nplus) \* \bigg[
            {151 \over 18} \* \S(1)
          + 2 \* \Ss(1,-2)
          - {11 \over 6} \* \S(2)
          \bigg] 
          \bigg)
  \nonumber\\&& \mbox{}
+ 4\, \* \cf \* \nf \* \bigg(
            {1 \over 12}
          + {4 \over 3} \* \S(1)
          - (\Nminus+\Nplus) \* \bigg[
            {11 \over 9} \* \S(1)
          - {1 \over 3} \* \S(2)
          \bigg]
          \bigg)
+ 4\, \* \cf^2 \* \bigg(
            4 \* \S(-3)
          + 2 \* \S(1)
          + 2 \* \S(2)  
          - {3 \over 8}
  \nonumber\\&& \mbox{}
          + \Nminus \* \bigg[
            \S(2)
          + 2 \* \S(3)
          \bigg]
          - (\Nminus+\Nplus) \* \bigg[
            \S(1)
          + 4 \* \Ss(1,-2)
          + 2 \* \Ss(1,2)
          + 2 \* \Ss(2,1)
          + \S(3)
          \bigg] 
          \bigg)
.\label{eq:gqq1}
\eea
At next-to-next-to-leading order (NNLO, N$^{\,2}$LO), we have 
re-calculated the coefficient function $c^{\, (2)}_{\, 2,\rm ns}$ of 
refs.~\cite{vanNeerven:1991nn,Moch:1999eb}, and we have computed all 
fermionic contributions to the splitting function $\gamma^{\,(2)}_{\,
\rm ns}$, the $n^1_{\! f}$ terms being a new result of this article. 
The fermionic NNLO corrections --- the complete expression for the 
$N$-space coefficient function can be found in ref.~\cite{Moch:1999eb} 
--- are given by
\bea
  && \! c^{\,(2)}_{\,2,\rm ns}(N) \: = \: 
 4\, \* \cf \* \nf \* \bigg(
           (1-\Nplus)  \*  \bigg[
            {122 \over 27} \* \S(1)
          + {7 \over 6} \* \Ss(1,1)
          \bigg]
          -  (\Nminus -1) \*  \bigg[
            {89 \over 108} \* \S(1)
          - \S(2)
          \bigg]
       - (\Nminus+\Nplus)  \* \bigg[
            {5 \over 6} \* \S(3)
  \nonumber\\&& \mbox{}
          + {13 \over 18} \* \Ss(1,1)
          + {1 \over 3} \* \Sss(1,1,1)
          - {2 \over 3} \* \Ss(2,1)
          - {1 \over 3} \* \Ss(1,2)
           \bigg] 
          - {1 \over 6} \* \Ss(1,1)
          + {457 \over 144}
          - {247 \over 108} \* \S(1)
          + {19 \over 6} \* \Nplus \* \S(2)
          \bigg)
,\label{eq:c2qq2}
\\[2mm]
  &&\gamma^{\,(2)}_{\,\rm ns}(N) \:\: = \:  
 16\, \* \ca \* \cf \* \nf \* \bigg( 
            {3 \over 2} \* \z3
          - {5 \over 4}
          + {10 \over 9} \* \S(-3)
          - {10 \over 9} \* \S(3)
          + {4 \over 3} \* \Ss(1,-2)
          - {2 \over 3} \* \S(-4)
          + 2 \* \Ss(1,1)
          - {25 \over 9} \* \S(2)
          + {257 \over 27} \* \S(1)
  \nonumber\\&& \mbox{}
          - {2 \over 3} \* \Ss(-3,1)
       - \Nplus \*  \bigg[
            \Ss(2,1)
          - {2 \over 3} \* \Ss(3,1)
          - {2 \over 3} \* \S(4)
          \bigg]
          + (1-\Nplus)  \*  \bigg[
            {23 \over 18} \* \S(3)
          - \S(2)
         \bigg]
       - (\Nminus+\Nplus)  \*   \bigg[
            \Ss(1,1)
          + {1237 \over 216} \* \S(1)
  \nonumber\\&& \mbox{}
          + {11 \over 18} \* \S(3)
          - {317 \over 108} \* \S(2)
          + {16 \over 9} \* \Ss(1,-2)
          - {2 \over 3} \* \Sss(1,-2,1)
          - {1 \over 3} \* \Ss(1,-3)
          - {1 \over 2} \* \Ss(1,3)
          - {1 \over 2} \* \Ss(2,1)
          - {1 \over 3} \* \Ss(2,-2)
          + \S(1) \* \z3
          + {1 \over 2} \* \Ss(3,1)
           \bigg]
          \bigg)
  \:\:\: \nonumber\\&& \mbox{}
+ 16\, \* \cf \* \nf^2 \* \bigg( 
            {17 \over 144}
          - {13 \over 27} \* \S(1)
          + {2 \over 9} \* \S(2)
              + (\Nminus+\Nplus) \*  \bigg[
            {2 \over 9} \* \S(1)
          - {11 \over 54} \* \S(2)
          + {1 \over 18} \* \S(3)
          \bigg]
          \bigg)
+  16\, \* \cf^2 \* \nf \* \bigg( 
            {23 \over 16}
          - {3 \over 2} \* \z3
  \nonumber\\&& \mbox{}
          + {4 \over 3} \* \Ss(-3,1)
          - {59 \over 36} \* \S(2)
          + {4 \over 3} \* \S(-4)
          - {20 \over 9} \* \S(-3)
          + {20 \over 9} \* \S(1)
          - {8 \over 3} \* \Ss(1,-2)
          - {8 \over 3} \* \Ss(1,1)
          - {4 \over 3} \* \Ss(1,2)
       + \Nplus \*  \bigg[
            {25 \over 9} \* \S(3)
          - {4 \over 3} \* \Ss(3,1)
  \nonumber\\&& \mbox{}
          - {1 \over 3} \* \S(4)
          \bigg]
               + (1-\Nplus)  \*  \bigg[
            {67 \over 36} \* \S(2)
          - {4 \over 3} \* \Ss(2,1)
          + {4 \over 3} \* \S(3)
          \bigg]
       + (\Nminus+\Nplus)  \*  \bigg[
            \S(1) \* \z3
          - {325 \over 144} \* \S(1)
          - {2 \over 3} \* \Ss(1,-3)
          + {32 \over 9} \* \Ss(1,-2)
  \nonumber\\&& \mbox{}
          - {4 \over 3} \* \Sss(1,-2,1)
          + {4 \over 3} \* \Ss(1,1)
          + {16 \over 9} \* \Ss(1,2)
          - {4 \over 3} \* \Ss(1,3)
          + {11 \over 18} \* \S(2)
          - {2 \over 3} \* \Ss(2,-2)
          + {10 \over 9} \* \Ss(2,1)
          + {1 \over 2} \* \S(4)
          - {2 \over 3} \* \Ss(2,2)
          - {8 \over 9} \* \S(3)
           \bigg]
          \bigg)
.\label{eq:gqq2}
\end{eqnarray}
  
The corresponding formulae for the longitudinal coefficient function
$C_{L,\rm ns}$ are deferred to the appendix, together with the rather 
lengthy $N$-space results for the fermionic parts of the third-order 
coefficient functions $c^{\,(3)}_{\,i,\rm ns}$, $i=2,L\,$, which 
partly also represent new results of this article (the $\nf^{\!\!2}$ 
term for $F_2$ has already been presented in ref.~\cite{Moch:2002yn}).
Notice that $c^{\,(3)}_{L,\rm ns}$ can be considered a NNLO quantity, 
since $C_L$ vanishes at order $\as^0$. On the other hand 
$c^{\,(3)}_{\,2,\rm ns}$ represents, at least at large $N$, the
dominant part of the N$^{\,3}$LO corrections to $F^{}_2$~\cite
{vanNeerven:2001pe}.

As briefly mentioned at the end of section \ref{sec:method}, we have
subjected our results to a number of checks. First of all, we have 
calculated some lower even moments in an arbitrary covariant gauge with 
the Mincer program~\cite{Gorishnii:1989gt,Larin:1991fz}, keeping the 
gauge parameter $\xi$ in the gluon propagator. All dependence on $\xi$ 
does cancel in the final results. 
Secondly the $\nf^{\!\!\!2}$-contribution to $\gamma_{\,\rm ns}^{\,(2)}$
is known from the work of Gracey \cite{Gracey:1994nn} and we agree 
with his result. Furthermore the coefficients of $\ln^{\,k} N$, 
$k = 3,\ldots 5$, of $c^{\,(3)}_{\,2,\rm ns}(N)$ agree with the 
prediction of the soft-gluon resummation \cite{Vogt:1999xa}.
Finally, we have checked the result of each individual diagram for 
several integer values of $N$ by comparing with the results of a Mincer 
calculation. Thus, as the strongest check, our results reproduce the 
fixed even moments $N=2,\dots,14$ computed in 
refs.~\cite{Larin:1994vu,Larin:1997wd,Retey:2000nq}. 
%
%
\section{Third-order results in {\bf x}-space}
\label{sec:xresults}
%
%
The $x$-space coefficient functions and the splitting functions are
obtained from the results of the previous section by an inverse Mellin
transformation, which maps the harmonic sums of moment space~\cite
{Gonzalez-Arroyo:1979df,Gonzalez-Arroyo:1980he,Vermaseren:1998uu,%
Blumlein:1998if} to harmonic polylogarithms in $x$-space~\cite
{Goncharov,Borwein,Remiddi:1999ew}. This transformation can be performed
by a completely algebraic procedure~\cite{Remiddi:1999ew,Moch:1999eb}
based on the fact that the harmonic sums also occur as coefficients
of the Taylor expansion of harmonic polylogarithms. 

\begin{figure}[thb]
\label{pic:Pqq2}
\centerline{\epsfig{file=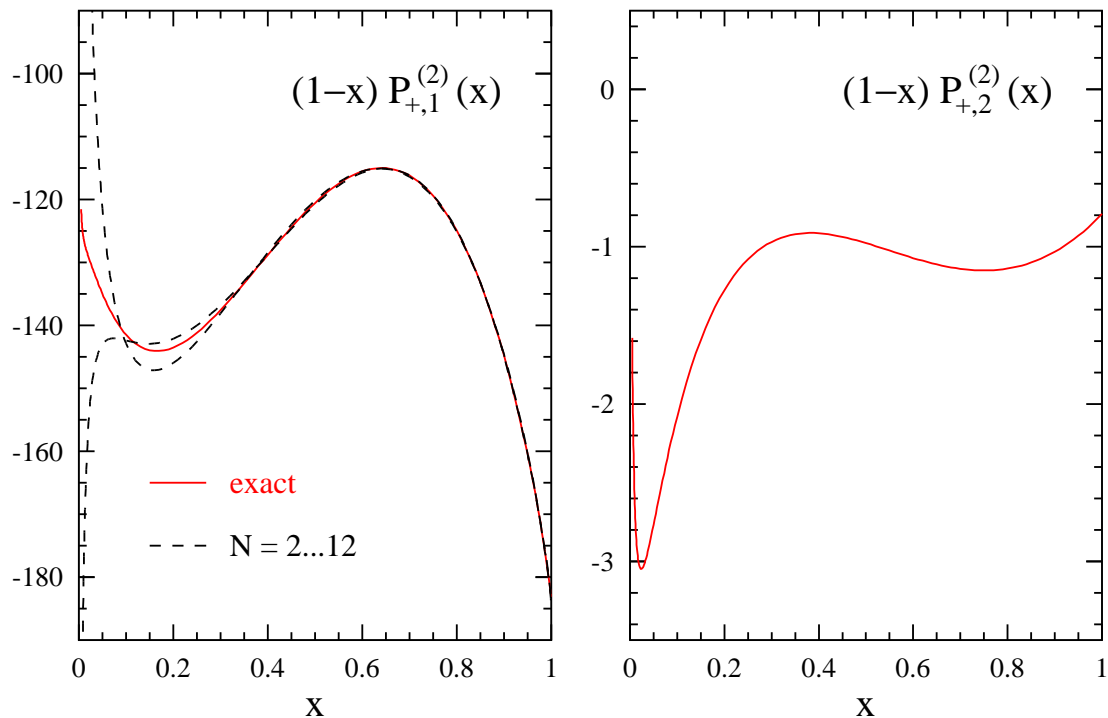,width=15cm,angle=0}}
\vspace{-1mm}
\caption{The $\nf^{\!\!1}$ and $\nf^{\!\!2}$ parts $P^{(2)}_{+,1}(x)$
 and $P^{(2)}_{+,2}(x)$ of the three-loop non-singlet splitting 
 function (\ref{eq:Pqq2}), multiplied by $(1\!-\!x)$ for display 
 purposes. Also shown in the left part (dashed curve) is the 
 uncertainty band derived in ref.~\cite{vanNeerven:2000wp} from the 
 lowest six even-integer moments~\cite{Larin:1994vu,Larin:1997wd,%
 Retey:2000nq}.}
\vspace{1mm}
\end{figure}

Here we confine ourselves to the third-order results; for the two-loop
non-singlet splitting functions and coefficient functions the reader is 
referred to refs.~\cite{Curci:1980uw,vanNeerven:1991nn,Moch:1999eb}.
For brevity the exact results are written down only for the splitting
function, conventionally related to the anomalous dimension 
(\ref{eq:gexp}) by
\beq
\label{eq:Pdef}
  \gamma^{\,(n)}_{\,\rm ns}(N) \: = \: - \int_0^1 \!dx\:\, x^{\,N-1}\, 
  P^{(n)}_{\rm ns}(x) \:\: .
\eeq
The fermionic part of $P^{(n)}_{\rm ns}$ involves only simpler harmonic 
polylogarithms which can be expressed in terms of the usual 
(poly-)logarithms. The $x$-space analogue of eq.~(\ref{eq:gqq2}),
graphically displayed in fig.~\ref{pic:Pqq2}, can thus be written as 
\bea
  &&P^{(2)}_{\rm ns}(x) \: = \:   
  16\, \* \ca \* \cf \* \nf \* \bigg(
         \pqq(x) \*  \bigg[
            {5 \over 9} \* \z2
          - {209 \over 216}
          - {3 \over 2} \* \z3
          - {167 \over 108} \* \ln(x)
          + {1 \over 3} \* \ln(x) \* \z2
          - {1 \over 4} \* \ln^2(x) \* \ln(1 - x) \:\:\:
  \nonumber\\&& \mbox{}
          - {7 \over 12} \* \ln^2(x)
          - {1 \over 18} \* \ln^3(x)
          - {1 \over 2} \* \ln(x)\* \Li(2,x)
          + {1 \over 3} \* \Li(3,x)
          \bigg]
       + \pqq( - x) \*  \bigg[
            {1 \over 2} \* \z3
          - {5 \over 9} \* \z2
          - {2 \over 3} \* \ln(1 + x) \* \z2
  \nonumber\\&& \mbox{}
          + {1 \over 6} \* \ln(x) \* \z2
          - {10 \over 9} \* \ln(x) \* \ln(1 + x)
          + {5 \over 18} \* \ln^2(x)
          - {1 \over 6} \* \ln^2(x) \* \ln(1 + x)
          + {1 \over 18} \* \ln^3(x)
          - {10 \over 9} \* \Li(2, - x)
  \nonumber\\&& \mbox{}
          - {1 \over 3} \* \Li(3, - x)
          - {1 \over 3} \* \Li(3,x)
          + {2 \over 3} \* \H(-1,0,1,x)
          \bigg]
       + (1+x) \*  \bigg[
            {1 \over 6} \* \z2
          + {1 \over 2} \* \ln(x)
          - {1 \over 2} \* \Li(2,x)
          - {2 \over 3} \* \Li(2, - x)
  \nonumber\\&& \mbox{}
          - {2 \over 3} \* \ln(x)\* \ln(1 + x)
          + {1 \over 24} \* \ln^2(x)
          \bigg]
       + (1-x) \*  \bigg[
            {1 \over 3} \* \z2
          - {257 \over 54}
          + \ln(1 - x)
          - {17 \over 9} \* \ln(x)
          - {1 \over 24} \* \ln^2(x)
          \bigg]
  \nonumber\\&& \mbox{}
       + \delta(1 - x) \* \bigg[
            {5 \over 4}
          - {167 \over 54} \* \z2
          + {1 \over 20} \* \z2^2
          + {25 \over 18} \* \z3
          \bigg]
          \bigg)
+ 16\, \* \cf \* \nf^2 \* \bigg(
         \pqq(x) \*  \bigg[
            {5 \over 54} \* \ln(x)
          - {1 \over 54}
          + {1 \over 36} \* \ln^2(x)
          \bigg]
  \nonumber\\&& \mbox{}
       + (1-x) \*  \bigg[
            {13 \over 54}
          + {1 \over 9} \* \ln(x)
          \bigg]
       - \delta(1 - x) \* \bigg[
            {17 \over 144}
          - {5 \over 27} \* \z2
          + {1 \over 9} \* \z3
          \bigg]
          \bigg)
+ 16\, \* \cf^2 \* \nf \*\bigg(
         \pqq(x) \*  \bigg[
            {5 \over 3} \* \z3
          - {55 \over 48}
  \nonumber\\&& \mbox{}
          + {5 \over 24} \* \ln(x)
          + {1 \over 3} \* \ln(x) \* \z2
          + {10 \over 9} \* \ln(x) \* \ln(1 - x)
          + {1 \over 4} \* \ln^2(x)
          + {2 \over 3} \* \ln^2(x) \* \ln(1 - x)
          + {2 \over 3} \* \ln(x) \* \Li(2,x)
  \nonumber\\&& \mbox{}
          - {2 \over 3} \* \Li(3,x)
          - {1 \over 18} \* \ln^3(x)
          \bigg]
       + \pqq( - x) \*  \bigg[
            {10 \over 9} \* \z2
          - \z3
          + {4 \over 3} \* \ln(1 + x) \* \z2
          - {1 \over 3} \* \ln(x) \* \z2
          - {5 \over 9} \* \ln^2(x)
  \nonumber\\&& \mbox{}
          + {20 \over 9} \* \ln(x)\* \ln(1 + x)
          - {1 \over 9}\* \ln^3(x)
          + {1 \over 3} \* \ln^2(x)\* \ln(1 + x)
          + {20 \over 9} \* \Li(2, - x)
          + {2 \over 3} \* \Li(3, - x)
          + {2 \over 3} \* \Li(3,x)
  \nonumber\\&& \mbox{}
          - {4 \over 3} \* \H(-1,0,1,x)
          \bigg]
       + (1+x) \*  \bigg[
            {7 \over 36} \* \ln^2(x)
          - {67 \over 72} \* \ln(x)
          + {4 \over 3} \* \ln(x)\* \ln(1 + x)
          + {1 \over 12} \* \ln^3(x)
          + {2 \over 3} \* \Li(2,x)
  \nonumber\\&& \mbox{}
          + {4 \over 3} \* \Li(2, - x)
          \bigg]
       + (1-x) \*  \bigg[
            {1 \over 9} \* \ln(x)
          - {10 \over 9}
          - {4 \over 3} \* \ln(1 - x)
          + {2 \over 3} \* \ln(x)\* \ln(1 - x)
          - {1 \over 3} \* \ln^2(x)
          \bigg]
  \nonumber\\&& \mbox{}
       - \delta(1 - x) \* \bigg[
            {23 \over 16}
          - {5 \over 12} \* \z2
          - {29 \over 30} \* \z2^2
          + {17 \over 6} \* \z3
          \bigg]
          \bigg)
\:\: ,\label{eq:Pqq2}
\eea
where we have introduced 
\beq
  p_{\rm{qq}}(x) \: = \: 2\, (1 - x)^{-1} - 1 - x
\eeq
and all divergences for $x \to 1 $ are understood in the sense of 
$+$-distributions. In eq.~(\ref{eq:Pqq2}) we have left one particular 
harmonic polylogarithm, $\H(-1,0,1,x)$, unsubstituted. This function is 
given by
\beq
  \H(-1,0,1,x) \: \equiv \: \int_0^x\, {dz \over 1+z}\: \Li(2,z) 
  \: = \: \Li(2,x) \ln(1+x) + \frac{1}{2} S_{1,2}(x^2) - S_{1,2}(-x)
          - S_{1,2}(x)\:\: , 
  \label{eq:hpoly}
\eeq
where the representation by the Nielsen function $S_{1,2}$ has been 
derived in ref.~\cite{Blumlein:2000hw}. $\H(-1,0,1,x)$ can also be 
expressed in terms of trilogarithms, albeit with more complicated 
arguments~\cite{Moch:1999eb}.

The $x$-space coefficient functions involve harmonic polylogarithms of
weight four, which in general cannot be expressed in terms of standard
polylogarithms and Nielsen functions anymore. Instead of writing down
the cumbersome exact expressions, we prefer to present sufficiently
accurate, compact parametrizations in terms of the $+$-distributions 
and end-point logarithms
\beq
\label{eq:logs}
  \DD_{\,k} \: = \:\left[ \frac{\ln^{\,k} (1-x)}{1-x} \right]_+
  \: ,\quad L_1 \: = \: \ln (1-x)
  \: ,\quad L_0 \: = \: \ln x \:\: .
\eeq
It is convenient to apply this procedure (which has been employed in 
ref.~\cite{vanNeerven:1999ca} for the two-loop coefficient functions) 
also to the $\nf^{\!\! 1}$ part of the splitting function 
(\ref{eq:Pqq2}). Inserting the numerical value of the QCD colour 
factors, this function can be approximated by 
\bea
\label{eq:Pappr}
  P^{(2)}_{\rm ns}(x)\!\! & \cong & \!\nf \:\big(
     - 183.187\: \DD_0 - 173.927\: \delta (1-x)
     - 5120/81\: L_1 - 197.0 + 381.1\: x + 72.94\: x^2
  \nonumber \\ & & \mbox{} \quad 
     + 44.79\: x^3 - 1.497\: xL_0^3 - 56.66\: L_0 L_1 
     - 152.6\: L_0 - 2608/81\: L_0^2 - 64/27\: L_0^3\, \big) 
  \nonumber \\[1mm] &+& \mbox{} \nf^{\!\!\! 2} \:\big(
     - \DD_0 - (51/16 + 3\,\z3 - 5\,\z2) \: \delta (1-x) 
     + x\,(1-x)^{-1} L_0\, (3/2\: L_0 + 5) + 1
  \nonumber \\ & & \mbox{} \quad
     + (1-x)\, (6 + 11/2\: L_0 + 3/4\: L_0^2)\, \big) \:\, 64/81 
  \:\: .
\eea
Corresponding parametrizations for the three-loop coefficient functions 
read
\bea
\label{eq:c2app}
  c_{\,2, \rm ns}^{\,(3)}(x)\!\! & \cong & \!\nf \:\big( \,
     640/81\: \DD_4 - 6592/81\: \DD_3 + 220.573\: \DD_2
     + 294.906\: \DD_1 - 729.359\: \DD_0 
  \nonumber \\ & & \mbox{} \quad
     + 2572.597\: \delta (1-x) 
     - 640/81\: L_1^4 + 167.2\: L_1^3 - 315.3\: L_1^2 + 4742\: L_1
  \nonumber \\ & & \mbox{} \quad
     + 762.1 + 7020\: x + 989.4\: x^2 
     + L_0 L_1\, (326.6 + 65.93\: L_0 + 1923\: L_1) 
  \nonumber \\ & & \mbox{} \quad
     + 260.1\: L_0 + 186.5\: L_0^2 + 12224/243\: L_0^3 
     + 728/243\: L_0^4 \, \big)
  \nonumber \\[1mm] &+& \mbox{} \nf^{\!\!\! 2} \:\big( \,
     64/81\: \DD_3 - 464/81\: \DD_2 + 7.67505\: \DD_1 + 1.00830\: \DD_0
     - 103.2655\: \delta (1-x)
  \nonumber \\ & & \mbox{} \quad
     - 64/81\: L_1^3 + 15.46\: L_1^2 - 51.71\: L_1
     + 59.00\: x + 70.66\: x^2 
     + L_0 L_1\, (- 80.05 
  \nonumber \\ & & \mbox{} \quad
       - 10.49\: L_0 + 41.67\: L_1)
     - 8.050\: L_0 - 1984/243\: L_0^2 - 368/243\: L_0^3 \, \big)
\:\: , \\[4mm] 
\label{eq:cLapp}
  c_{\,L, \rm ns}^{\,(3)}(x)\!\! & \cong & \!\nf \:\big( \,
     1024/81\: L_1^3 - 112.4\: L_1^2 + 340.3\: L_1
     + 409 - 210\: x - 762.6\: x^2 - 1792/81\: x L_0^3
  \nonumber \\ & & \mbox{} \quad
     + L_0 L_1\, (969.2 + 304.8\: L_0 - 288.2\: L_1)
     + 200.8\: L_0 + 64/3\: L_0^2 + 0.046\: \delta (1-x) \, \big)
  \nonumber \\[1mm] &+& \mbox{} \nf^{\!\!\! 2} \:\big( \,
     3x\: L_1^2 + (6 - 25\: x)\, L_1
     - 19 + (317/6 - 12\, \z2) \, x
     - 6x \: L_0 L_1 + 6x\: \Li(2,x) 
  \nonumber \\ & & \mbox{} \quad
     + 9x \:  L_0^2 - (6 - 50\: x) \, L_0 \,) \:\, 64/81
\:\: .
\eea
The $\nf^{\!\!2}$ parts of $P^{(2)}_{\rm ns}$ and 
$c_{\,L,\rm ns}^{\,(3)}$, the $+$-distribution contributions (up to a 
numerical truncation of the coefficients involving $\zeta_{i\,}$), and 
the rational coefficients of the (sub-)leading regular end-point terms 
are exact in eqs.~(\ref{eq:Pappr}) -- (\ref{eq:cLapp}).
The remaining coefficients have been determined by fits to the exact 
results, for which we have used the Fortran package of 
ref.~\cite{Gehrmann:2001pz}. The above parametrizations deviate from 
the exact expressions by one part in thousand or less, an accuracy 
which should be amply sufficient for foreseeable numerical applications.
%
%
\section{Implications for the threshold resummation}
\label{sec:sresults}
%
%
The large-$N\,$/$\,$large-$x$ behaviour of the three-loop splitting 
functions and coefficient functions is of special interest in 
connection with the soft-gluon (threshold) exponentiation~\cite
{Sterman:1987aj,Catani:1989ne,Catani:1991rp} at next-to-next-leading
logarithmic (NNL) accuracy. Here the coefficient function for 
$F^{}_{2,\rm ns}$ can, up to terms which vanish for $N \to \infty$, be 
written as
\beq
\label{eq:csoft}
  C_{\,2,\rm ns}(\as, N) \: =\: 
  (1 + a_{\rm s\,} g_{01}^{} + a_{\rm s\,}^2 g_{02}^{} + \ldots ) \: 
  \exp \, [\, L\, g_1^{}(a_{\rm s\,}L) + g_2^{}(a_{\rm s\,}L) + 
  a_{\rm s}\, g_3^{}(a_{\rm s\,}L) + \ldots\, ] 
\eeq
with $a_{\rm s} = \as/(4\pi)$ and $L = \ln N$. The functions 
$g_l^{}$ depend on (universal) coefficients $A_{\,i\,\leq\, l}$ and 
$B_{\,i\,\leq\, l\!-\!1}$ and process-dependent parameters 
$D_{\,i\,\leq\, l\!-\!1}^{\,\rm DIS}$ as described in 
ref.~\cite{Vogt:2000ci}, where also the explicit expressions for the 
functions $g_{1,2,3}^{}$ can be found. Hence the NNL function $g_3^{}$
involves the new coefficients $A_3$, $B_2$ and $D_2^{\,\rm DIS}$. These 
coefficients can be fixed by expanding eq.~(\ref{eq:csoft}) in powers 
of $\as$ and comparing to the result of the full fixed-order 
calculations.
 
In the \MSb\ scheme adopted in this article, the parameter $A_3$ is 
simply the coefficient of $\ln N$ in $\gamma_{\,\rm ns}^{(2)}(N)$ 
or, equivalently, of $1/(1-x)_+$ in $P_{\,\rm ns}^{(2)}(x)$. Its 
fermionic part is thus known from eq.~(\ref{eq:Pqq2}), 
\beq
   A_3\Big|_{\,\nf} \: = \:
   C_A C_F \nf\, \left[ - \frac{836}{27} + \frac{160}{9}\:\z2 
               - \frac{112}{3}\:\z3 \right]
   + C_F^{\,2} \nf\, \left[ -  \frac{110}{3}  + 32\:\z3 \right]
   + C_F \nf^{\!\!2} \left[ - \frac{16}{27}\,\right] \:\: .
\eeq
The numerical value can be read off from eq.~(\ref{eq:Pappr}). Like for
the whole of $P_{\,\rm ns}^{(2)}(x)$, as shown in fig.~\ref{pic:Pqq2},
this result is consistent with, but supersedes the estimate derived
in ref.~\cite{vanNeerven:2000wp} from the first six even-integer 
moments. Parallel to our work $A_3|_{\,\nf}$ has also been calculated
in ref.~\cite{cberger}.

The combination $B_2 + D_2^{\,\rm DIS}$ has been determined in ref.\
\cite{Vogt:2000ci} by comparing the expansion of eq.~(\ref{eq:csoft})
to the $\ln N$ term of the two-loop coefficient function 
$c_{\,2,\rm ns}^{\,(2)}$ of ref.~\cite{vanNeerven:1991nn}. As the 
$\ln^2 N$ (or $\DD_1$) contribution to $c_{\,2,\rm ns}^{\,(3)}$ involves
a different linear combination, $\beta_0 (B_2 + 2\,D_2^{\,\rm DIS})$,
of the very same coefficients, $B_2$ and $D_2^{\,\rm DIS}$ can be 
disentangled using the three-loop coefficient function. The analytic 
results for the two new $+$-distribution coefficients read
\bea
\label{eq:c2d1}
  c_{\,2,\rm ns}^{\,(3)} \Big|_{\,\DD^{}_1 \nf} \! & = \! &
  C_A C_F \nf\, \left[ - \frac{15062}{81} + \frac{512}{9}\:\z2
                + 16\:\z3 \right] \: + \:
  C_F^{\,2} \nf\, \left[ \frac{83}{9} + 168\:\z2 + 
                \frac{112}{3}\:\z3 \right]
  \nonumber \\ & & \mbox{} \! + \:
  C_F \nf^{\!\!2} \left[ \frac{940}{81} - \frac{32}{9}\:\z2 \right]
  \:\: ,
  \\[2mm] 
  c_{\,2,\rm ns}^{\,(3)} \Big|_{\,\DD^{}_0 \nf} \! & = \! &
  C_A C_F \nf\, \left[ - \frac{160906}{729} - \frac{9920}{81}\:\z2
    - \frac{776}{9}\:\z3 + \frac{208}{15}\:\z2^{\!\! 2}\right] \: + \:
  C_F^{\,2} \nf\, \left[ - \frac{2003}{108} \right.
  \nonumber \\ & & \mbox{} 
   \left. - \frac{4226}{27}\:\z2
    - 60\:\z3 + 16\:\z2^{\!\! 2}\right] \: + \
  C_F \nf^{\!\!2} \left[ - \frac{8714}{729} + \frac{232}{27}\:\z2 
    - \frac{32}{27}\:\z3 \right] 
\label{eq:c2d2}
\eea
(the coefficients of $\DD^{}_{2,\ldots, 5}$ can be found in ref.~\cite
{Vogt:1999xa}). In fact, due to the prefactor $\beta_{0\,}$, the 
complete results for $B_2$ and $D_2^{\,\rm DIS}$ can already be 
inferred from fermionic result (\ref{eq:c2d1}), yielding
\bea
\label{eq:B2D2}
  B_2 \: & =\! &
  C_F^{\,2}  \left[ -\frac{3}{2} - 24\,\zeta_3 + 12\,\zeta_2 \right] 
  \: +\: C_F C_A  \left[ - \frac{3155}{54} + 40\,\zeta_3
                 + \frac{44}{3}\,\zeta_2 \right] 
  \nonumber \\ & & \mbox{} + \:
  C_F \nf\,  \left[ \frac{247}{27} - \frac{8}{3}\,\zeta_2 \right] 
  \:\: , \\
  D_2^{\,\rm DIS} \!\! & =\! & 0 \:\: .
\eea
The vanishing of $D_1^{\,\rm DIS}$ and $D_2^{\,\rm DIS}$ --- in contrast
to the Drell-Yan process, where $D_2$ is different from zero~\cite
{Vogt:2000ci} --- calls for a deeper explanation, possibly offering
an all-order generalization. 

Finally we note that, once the non-fermionic contributions to the
3-loop non-singlet splitting functions and coefficient functions
are completed, the NNL threshold resummation facilitates a prediction
of the first six towers of logarithms, i.e., the coefficients of
$\as^{\,n} \ln^{\,2n-i} N$, $i = 0, \ldots , 5$, of $C_{\,2,\rm ns}$ at 
all orders $n>3$. We will return to this issue in a later publication. 
%
%
\section{Summary}
\label{sec:summary}
%
%
We have computed the fermionic ($\nf$-enhanced) third-order 
contributions to the structure functions $F_2$ and $F_L$ in 
electromagnetic deep-inelastic scattering. The calculation has been
carried out for all even-integer Mellin moments $N$, by solving the
three-loop integrals by means of recursion relations (difference
equations) in $N$. This progress with respect to previous computations
restricted to some fixed moments $N$ is especially due to an improved
understanding of the mathematics of harmonic sums and difference
equations, and the implementation of corresponding tools in the
symbolic manipulation program FORM which we employed to handle the
huge intermediate expressions. We are confident that our approach will
enable us to compute all three-loop corrections in~DIS.

We have thus been able to derive the complete expressions for the
corresponding $\nf$-parts of the NNLO anomalous dimensions and splitting
functions and the N$^3$LO coefficient functions for $F_2$ and $F_L$.
The results have been presented in both Mellin-$N$ and Bjorken-$x$ 
space, in the latter case we have also provided easy-to-use accurate 
parametrizations. Our results agree with all partial and approximate
results available in the literature for these quantities, in particular
we reproduce the even-integer moments $N=2,\ldots,12$ computed before.

The present results for the three-loop splitting function represent 
a step towards completing the ingredients required for NNLO 
calculations of hard-scattering processes involving initial-state
hadrons in perturbative QCD. The three-loop coefficient functions for
the most important structure function $F_2$ form the dominant part of 
the N$^3$LO corrections at large $x$, thus facilitating extractions of
$\as$ with a distinctly reduced theoretical uncertainty. Already the 
$\nf$-part computed in this article leads to a complete determination 
of the threshold-resummation parameters $B_2$ and $D_2^{\,\rm DIS}$ 
--- including the non-fermionic contributions --- of which only the sum 
was known so far, thus practically completing the information required 
for the next-to-next-to-leading logarithmic resummation. 
\subsection*{Acknowledgments}
The work of S.M. at the {\em Institut f{\"u}r Theoretische 
Teilchenphysik} at Karlsruhe University was supported by the German 
Research Society (DFG) under contract No.~FOR 264/2-1. The work of 
J.V. and A.V. is part of the research program of the Dutch Foundation 
for Fundamental Research of Matter (FOM). 
%
%
\setcounter{equation}{0}
\renewcommand{\theequation}{A.\arabic{equation}}
\begin{appendix}
\section{Appendix \label{sec:appendix}}
%
%
All results for the non-singlet anomalous dimensions and coefficient 
functions presented in this article can be obtained as a FORM file from 
the preprint server {\tt http://arXiv.org } by downloading the source 
file. Furthermore they are available from the authors upon request.

The fermionic parts, i.e., all terms proportional to $C_A C_F \nf$, 
$C_F^{\,2} \nf$ and $C_F n_{\!f}^{\,2}$ of the three-loop coefficient 
function for the electromagnetic structure function $F_2$ are given by
\bea
  &&c^{(3)}_{2,\rm ns}(N) \: = \: 
 16\: \* \ca \* \cf \* \nf \*  \bigg(
  \delta(N\minus 2) \* \bigg[
            {5 \over 3} \* \z5
          - {119 \over 300} \* \z3
          \bigg]
          + {142883 \over 7776}
          - {1051 \over 72} \* \z3
          + {3 \over 4} \* \z4
          + {83 \over 54} \* \S(4)
          - 2 \* \Ss(1,4)
  \quad \nonumber\\&&
          - {4 \over 9} \* \Ss(-4,1)
          + {191 \over 81} \* \S(-3)
          - {16 \over 3} \* \Ss(-3,-2)
          + {20 \over 27} \* \Ss(-3,1)
          - {4 \over 9} \* \Sss(-3,1,1)
          - {29 \over 18} \* \S(-2)
          - {14 \over 3} \* \S(-2) \* \z3
          - {16 \over 3} \* \Ss(-2,-3)
  \nonumber\\&&
          + {13 \over 3} \* \Ss(-2,-2)
          + {8 \over 3} \* \Sss(-2,-2,1)
          + {23 \over 9} \* \S(-5)
          + {199819 \over 8100} \* \S(1)
          - {101 \over 18} \* \Ss(1,2)
          + {181 \over 108} \* \Ss(3,1)
          - {83 \over 54} \* \S(-4)
          + {4132 \over 135} \* \Ss(1,-2)
  \nonumber\\&&
          + {8 \over 3} \* \Sss(1,-2,-2)
          + {56 \over 9} \* \Sss(1,-2,1)
          + {21463 \over 1080} \* \Ss(1,1)
          - {10 \over 3} \* \Ss(1,1) \* \z3
          - 4 \* \Sss(1,1,-3)
          + {58 \over 9} \* \Sss(1,1,1)
          - {23 \over 18} \* \Sss(1,2,1)
          + 10 \* \Ss(1,-4)
  \nonumber\\&&
          + {8 \over 3} \* \Ssss(1,1,1,-2)
          + {1 \over 3} \* \Ssss(1,1,1,2)
          - 4 \* \Sss(2,-2,1)
          - {1 \over 3} \* \Ssss(1,1,2,1)
          - {8 \over 3} \* \Sss(1,2,-2)
          + \Sss(1,2,2)
          + {35 \over 9} \* \Ss(1,3)
          + {5 \over 3} \* \Sss(1,3,1)
          - {1 \over 12} \* \Ss(2,2)
  \nonumber\\&&
          - {36719 \over 16200} \* \S(2)
          + {32 \over 3} \* \S(2) \* \z3
          + 10 \* \Ss(2,-3)
          - {218 \over 9} \* \Ss(2,-2)
          + {23 \over 18} \* \Sss(1,1,2)
          - {263 \over 60} \* \Ss(2,1)
          - {8 \over 3} \* \Sss(2,1,-2)
          - 25 \* \S(1) \* \z3 
  \nonumber\\&&
          - {2 \over 3} \* \Ss(2,3)
          - {4537 \over 1620} \* \S(3)
          + {28 \over 3} \* \Ss(3,-2)
          - {208 \over 9} \* \Ss(1,-3)
          - {112 \over 9} \* \Nminus \* \Sss(1,1,-2)
      + \Nplus \* \bigg[
            {4 \over 9} \* \Sss(3,1,1)
          - {5 \over 3} \* \Sss(2,1,1)
          + {4 \over 9} \* \Ss(4,1)
  \nonumber\\&&
          - {23 \over 9} \* \S(5)
          \bigg]
      + (\Nminusthree-\Nminustwo) \* \bigg[ 
            {2 \over 5} \* \S(1) \* \z3
          + {2 \over 5} \* \Ss(1,-3)
          - {119 \over 450} \* \Ss(1,-2)
          - {2 \over 15} \* \Sss(1,-2,1)
          - {2 \over 3} \* \Ss(1,1) \* \z3
          - {2 \over 15} \* \Sss(1,1,-2)
  \nonumber\\&&
          + {1 \over 3} \* \Sss(1,1,3)
          - {1 \over 3} \* \Sss(1,3,1)
          + {4 \over 15} \* \Ss(2,-2)
          \bigg]
      + (\Nminustwo-\Nminus) \* \bigg[ 
            {2 \over 5} \* \S(3)
          - {179 \over 450} \* \S(1)
          + {2 \over 3} \* \S(1) \* \z3
          + {1 \over 5} \* \Ss(1,1)
          - {59 \over 450} \* \S(2)
  \nonumber\\&&
          + {1 \over 5} \* \Ss(2,1)
          - {1 \over 3} \* \Ss(1,3)
          + {2 \over 15} \* \Ss(1,-2)
          + {1 \over 3} \* \Ss(3,1)
          \bigg]
      + (\Nminus-1) \* \bigg[ 
            4 \* \Ss(1,-4)
          + 2 \* \Sss(1,-2,-2)
          + {2 \over 3} \* \Sss(1,2,2)
          \bigg]
  \nonumber\\&&
      + (1-\Nplus) \* \bigg[ 
            {11057 \over 324} \* \S(1)
          - {21 \over 2} \* \S(1) \* \z3
          + {4 \over 9} \* \Ss(1,-3)
          - {7 \over 3} \* \Ss(1,-2)
          - {14 \over 3} \* \Sss(1,-2,-2)
          - {8 \over 3} \* \Sss(1,-2,1)
          + {3559 \over 216} \* \Ss(1,1)
  \nonumber\\&&
          - 8 \* \Ss(1,1) \* \z3
          - 8 \* \Sss(1,1,-3)
          + {176 \over 9} \* \Sss(1,1,-2)
          + {217 \over 36} \* \Sss(1,1,1)
          + {16 \over 3} \* \Ssss(1,1,1,-2)
          + {2 \over 3} \* \Ssss(1,1,1,2)
          - {7 \over 36} \* \Sss(1,1,2)
  \nonumber\\&&
          - {2 \over 3} \* \Ssss(1,1,2,1)
          + {2 \over 3} \* \Sss(1,1,3)
          - {55 \over 9} \* \Ss(1,2)
          - {16 \over 3} \* \Sss(1,2,-2)
          + {7 \over 36} \* \Sss(1,2,1)
          - {4 \over 3} \* \Sss(1,2,2)
          + {38 \over 9} \* \Ss(1,3)
          + {8 \over 3} \* \Sss(1,3,1)
  \nonumber\\&&
          - 4 \* \Ss(1,4)
          - {231037 \over 5400} \* \S(2)
          + 3 \* \S(2) \* \z3
          - 6 \* \Ss(2,-3)
          + {118 \over 9} \* \Ss(2,-2)
          + {20 \over 3} \* \Sss(2,-2,1)
          - {793 \over 90} \* \Ss(2,1)
          - {16 \over 3} \* \Sss(2,1,-2)
  \nonumber\\&&
          + {1 \over 6} \* \Sss(2,1,2)
          - {4 \over 9} \* \Ss(2,2)
          - {1 \over 6} \* \Sss(2,2,1)
          - 2 \* \Ss(2,3)
          + {49717 \over 1620} \* \S(3)
          - {22 \over 3} \* \Ss(3,-2)
          - {47 \over 54} \* \Ss(3,1)
          - {1 \over 6} \* \Ss(3,2)
          - {166 \over 27} \* \S(4)
  \nonumber\\&&
          + {5 \over 6} \* \Ss(4,1)
          \bigg]
      + (\Nplus-\Nplustwo) \* \bigg[ 
            10 \* \S(1) \* \z3
          - {219 \over 50} \* \S(1)
          + {6 \over 5} \* \Ss(1,-2)
          + 4 \* \Sss(1,-2,1)
          + {24 \over 5} \* \Ss(1,1)
          + 4 \* \Ss(1,1) \* \z3
  \nonumber\\&&
          - 4 \* \Sss(1,1,-2)
          - 2 \* \Sss(1,1,3)
          + 3 \* \Ss(1,2)
          - 3 \* \Ss(1,3)
          + 2 \* \Sss(1,3,1)
          - {21 \over 50} \* \S(2)
          - 4 \* \S(2) \* \z3
          + 4 \* \Ss(2,-2)
          - {24 \over 5} \* \Ss(2,1)
          + 2 \* \Ss(2,3)
  \nonumber\\&&
          - {3 \over 5} \* \S(3)
          - \Ss(3,1)
          - 2 \* \Ss(4,1)
          \bigg]
      + (\Nplustwo-\Nplusthree) \* \bigg[ 
            3 \* \Ss(2,2)
          - 3 \* \Sss(1,1,2)
          - {72 \over 5} \* \S(1) \* \z3
          + {18 \over 5} \* \Ss(1,-3)
          - {159 \over 50} \* \Ss(1,-2)
  \nonumber\\&&
          - {6 \over 5} \* \Sss(1,-2,1)
          - 6 \* \Ss(1,1) \* \z3
          - {6 \over 5} \* \Sss(1,1,-2)
          + 3 \* \Sss(1,1,3)
          + 3 \* \Sss(1,2,1)
          + 6 \* \S(2) \* \z3
          - 3 \* \Sss(1,3,1)
          + {6 \over 5} \* \Ss(2,-2)
          + 3 \* \Ss(4,1)
  \nonumber\\&&
          - 3 \* \Ss(2,3)
          - {159 \over 50} \* \S(3)
          - {21 \over 5} \* \Ss(3,1)
          + {18 \over 5} \* \S(4)
          \bigg]
       + (\Nminus+\Nplus) \* \bigg[
            {1711 \over 108} \* \S(1) \* \z3
          - {5608067 \over 291600} \* \S(1)
          - {1 \over 2} \* \S(1) \* \z4
  \nonumber\\&&
          - {79 \over 9} \* \Ss(1,-4)
          + 4 \* \Sss(1,1,-2)
          + {392 \over 27} \* \Ss(1,-3)
          + {2 \over 9} \* \Sss(1,-3,1)
          - {6613 \over 405} \* \Ss(1,-2)
          - {104 \over 27} \* \Sss(1,-2,1)
          + {4 \over 9} \* \Ssss(1,-2,1,1)
  \nonumber\\&&
          + {8 \over 3} \* \Ss(1,1) \* \z3
          - {25511 \over 1620} \* \Ss(1,1)
          + {52 \over 9} \* \Sss(1,1,-3)
          - {184 \over 27} \* \Sss(1,1,1)
          - {32 \over 9} \* \Ssss(1,1,1,-2)
          - {11 \over 9} \* \Ssss(1,1,1,1)
          - {17 \over 18} \* \Ssss(1,1,1,2)
  \nonumber\\&&
          + {1 \over 36} \* \Sss(1,1,2)
          + {4 \over 9} \* \Ssss(1,1,2,1)
          + {13 \over 18} \* \Sss(1,1,3)
          + {661 \over 108} \* \Ss(1,2)
          + {34 \over 9} \* \Sss(1,2,-2)
          + {29 \over 12} \* \Sss(1,2,1)
          + {1 \over 2} \* \Ssss(1,2,1,1)
          - 4 \* \Ss(1,3)
          + \Ss(4,1)
  \nonumber\\&&
          - {8 \over 3} \* \Sss(1,3,1)
          + {17 \over 9} \* \Ss(1,4)
          + {44537 \over 2700} \* \S(2)
          - {23 \over 3} \* \S(2) \* \z3
          - {79 \over 9} \* \Ss(2,-3)
          + {377 \over 27} \* \Ss(2,-2)
          + {20 \over 9} \* \Sss(2,-2,1)
          + {5731 \over 540} \* \Ss(2,1)
  \nonumber\\&&
          + {34 \over 9} \* \Sss(2,1,-2)
          + {59 \over 18} \* \Sss(2,1,1)
          + {5 \over 9} \* \Sss(2,1,2)
          - {53 \over 18} \* \Ss(2,2)
          - {5 \over 9} \* \Sss(2,2,1)
          + {13 \over 18} \* \Ss(2,3)
          - {4511 \over 405} \* \S(3)
          - {67 \over 9} \* \Ss(3,-2)
          - {83 \over 18} \* \Ss(3,1)
  \nonumber\\&&
          - {5 \over 6} \* \Sss(3,1,1)
          + {1 \over 2} \* \Ss(3,2)
          + {253 \over 54} \* \S(4)
          \bigg]
          \bigg)
+ 16\: \* \cf \* \nf^2 \* \bigg( 
            {1 \over 18} \* \Sss(1,1,1)
          - {9517 \over 7776}
          - {1 \over 18} \* \z3
          - {757 \over 648} \* \S(1)
          - {29 \over 108} \* \Ss(1,1)
  \nonumber\\&&
          - {1 \over 18} \* \Ss(1,2)
          - {43 \over 324} \* \S(2)
          - {1 \over 6} \* \Ss(2,1)
          + {19 \over 54} \* \S(3)
      + (1-\Nplus) \* \bigg[
            {13 \over 18} \* \Ss(2,1)
          - {19 \over 18} \* \S(3)
          + {265 \over 108} \* \S(2)
          + {7 \over 18} \* \Ss(1,2)
          - {7 \over 18} \* \Sss(1,1,1)
  \nonumber\\&&
          - {133 \over 108} \* \Ss(1,1)
          - {1421 \over 648} \* \S(1)
          \bigg]
       + (\Nminus+\Nplus)  \*  \bigg[
            {5585 \over 5832} \* \S(1)
          + {1 \over 27} \* \S(1) \* \z3
          + {161 \over 324} \* \Ss(1,1)
          + {13 \over 54} \* \Sss(1,1,1)
          + {1 \over 9} \* \Ssss(1,1,1,1)
  \nonumber\\&&
          - {1 \over 9} \* \Sss(1,1,2)
          - {13 \over 54} \* \Ss(1,2)
          - {1 \over 9} \* \Sss(1,2,1)
          + {1 \over 9} \* \Ss(1,3)
          - {301 \over 324} \* \S(2)
          - {29 \over 54} \* \Ss(2,1)
          - {2 \over 9} \* \Sss(2,1,1)
          + {2 \over 9} \* \Ss(2,2)
          + {62 \over 81} \* \S(3)
          + {1 \over 3} \* \Ss(3,1)
  \nonumber\\&&
          - {23 \over 54} \* \S(4)
          \bigg]
          \bigg)
+ 16\: \* \cf^2 \* \nf \* \bigg(
            {119 \over 150} \* \z3 \* \delta(N\minus 2) 
          - {341 \over 576}
          + {139 \over 6} \* \z3
          - {3 \over 4} \* \z4
          - {46 \over 9} \* \S(-5)
          + {8 \over 9} \* \Ss(-4,1)
          - {382 \over 81} \* \S(-3)
  \nonumber\\&&
          + {83 \over 27} \* \S(-4)
          + {32 \over 3} \* \Ss(-3,-2)
          - {40 \over 27} \* \Ss(-3,1)
          + {8 \over 9} \* \Sss(-3,1,1)
          + {29 \over 9} \* \S(-2)
          + {28 \over 3} \* \S(-2) \* \z3
          - {16 \over 3} \* \Sss(-2,-2,1)
          + {66367 \over 129600} \* \S(1)
  \nonumber\\&&
          + {32 \over 3} \* \Ss(-2,-3)
          - {26 \over 3} \* \Ss(-2,-2)
          + {301 \over 6} \* \S(1) \* \z3
          - 4 \* \Ss(1,-4)
          + {416 \over 9} \* \Ss(1,-3)
          - {8264 \over 135} \* \Ss(1,-2)
          + {8 \over 3} \* \Sss(1,-2,-2)
  \nonumber\\&&
          - {112 \over 9} \* \Sss(1,-2,1)
          + {83 \over 144} \* \Ss(1,1)
          + {20 \over 3} \* \Ss(1,1) \* \z3
          + 8 \* \Sss(1,1,-3)
          - {224 \over 9} \* \Sss(1,1,-2)
          - {557 \over 72} \* \Sss(1,1,1)
          - {16 \over 3} \* \Ssss(1,1,1,-2)
  \nonumber\\&&
          - {2 \over 3} \* \Ssss(1,1,1,2)
          - {31 \over 18} \* \Sss(1,1,2)
          + {2171 \over 216} \* \Ss(1,2)
          + {16 \over 3} \* \Sss(1,2,-2)
          + {7 \over 3} \* \Sss(1,2,1)
          + {2 \over 3} \* \Sss(1,2,2)
          - {235 \over 36} \* \Ss(1,3)
          + 4 \* \Ss(1,4)
  \nonumber\\&&
          - {10 \over 3} \* \Sss(1,3,1)
          - {140237 \over 16200} \* \S(2)
          - 22 \* \S(2) \* \z3
          - 20 \* \Ss(2,-3)
          + {436 \over 9} \* \Ss(2,-2)
          + 8 \* \Sss(2,-2,1)
          + {7133 \over 1080} \* \Ss(2,1)
          + {16 \over 3} \* \Sss(2,1,-2)
  \nonumber\\&&
          + {50 \over 9} \* \Sss(2,1,1)
          - 4 \* \Ss(2,2)
          + {7 \over 3} \* \Ss(2,3)
          - {7627 \over 3240} \* \S(3)
      + \Nplus  \*  \bigg[
            \Ssss(2,1,1,1)
          - \Sss(2,1,2)
          + {2 \over 3} \* \Ssss(1,1,2,1)
          - \Sss(2,2,1)
          - {56 \over 3} \* \Ss(3,-2)
  \nonumber\\&&
          - {329 \over 27} \* \Ss(3,1)
          - {29 \over 9} \* \Sss(3,1,1)
          + {7 \over 3} \* \Ss(3,2)
          + {401 \over 54} \* \S(4)
          + {53 \over 18} \* \Ss(4,1)
          + {1 \over 18} \* \S(5)
          \bigg]
      + (\Nminusthree-\Nminustwo)  \*  \bigg[
            {119 \over 225} \* \Ss(1,-2)
  \nonumber\\&&
          - {4 \over 5} \* \S(1) \* \z3
          + {4 \over 15} \* \Sss(1,-2,1)
          - {4 \over 5} \* \Ss(1,-3)
          + {4 \over 15} \* \Sss(1,1,-2)
          - {8 \over 15} \* \Ss(2,-2)
          \bigg]
      + (\Nminustwo-\Nminus)   \*  \bigg[
            {179 \over 225} \* \S(1)
          - {4 \over 15} \* \Ss(1,-2)
  \nonumber\\&&
          + {4 \over 15} \* \Ss(1,1)
          + {59 \over 225} \* \S(2)
          + {4 \over 15} \* \Ss(2,1)
          - {4 \over 5} \* \S(3)
          \bigg]
          + {15439 \over 1440} \* (\Nminus - 1)  \* \Ss(1,1)
      + (\Nplus-\Nplustwo)  \*  \bigg[
            {219 \over 25} \* \S(1)
          - {12 \over 5} \* \Ss(1,-2)
  \nonumber\\&&
          + 4 \* \S(1) \* \z3
          - 8 \* \Sss(1,-2,1)
          - {13 \over 5} \* \Ss(1,1)
          + 8 \* \Sss(1,1,-2)
          + 2 \* \Sss(1,1,2)
          - 5 \* \Ss(1,2)
          - 2 \* \Sss(1,2,1)
          - {154 \over 25} \* \S(2)
          + {36 \over 5} \* \S(3)
          + 10 \* \Ss(3,1)
  \nonumber\\&&
          + {13 \over 5} \* \Ss(2,1)
          - 2 \* \Ss(2,2)
          - 8 \* \Ss(2,-2)
          \bigg]
      + (\Nplustwo-\Nplusthree)  \*  \bigg[
            5 \* \Sss(1,1,2)
          - {12 \over 5} \* \Ss(2,-2) 
          - 5 \* \Ss(2,2)
          + {37 \over 5} \* \Ss(3,1)
          + {159 \over 25} \* \Ss(1,-2)
  \nonumber\\&&
          + {159 \over 25} \* \S(3)
          - 5 \* \Sss(1,2,1)
          + {12 \over 5} \* \Sss(1,1,-2)
          + {12 \over 5} \* \Sss(1,-2,1)
          + {114 \over 5} \* \S(1) \* \z3
          - {36 \over 5} \* \Ss(1,-3)
          - {36 \over 5} \* \S(4)
          \bigg]
      + (1-\Nplus)  \*  \bigg[
           4 \* \S(4)
  \nonumber\\&&
          - {10337 \over 576} \* \S(1)
          - 8 \* \Ss(1,-4)
          - {8 \over 9} \* \Ss(1,-3)
          + {14 \over 3} \* \Ss(1,-2)
          + {16 \over 3} \* \Sss(1,-2,-2)
          + {16 \over 3} \* \Sss(1,-2,1)
          + {40 \over 3} \* \Ss(1,1) \* \z3
          + 16 \* \Sss(1,1,-3)
  \nonumber\\&&
          - {76157 \over 4320} \* \Ss(1,1)
          - {128 \over 9} \* \Sss(1,1,-2)
          + {179 \over 72} \* \Sss(1,1,1)
          - {32 \over 3} \* \Ssss(1,1,1,-2)
          + {14 \over 3} \* \Ssss(1,1,1,1)
          - {4 \over 3} \* \Ssss(1,1,1,2)
          + {565897 \over 16200} \* \S(2)
  \nonumber\\&&
          + 2 \* \Ssss(1,1,2,1)
          - {295 \over 72} \* \Ss(1,2)
          + {32 \over 3} \* \Sss(1,2,-2)
          - {17 \over 3} \* \Sss(1,2,1)
          + {4 \over 3} \* \Sss(1,2,2)
          - {115 \over 36} \* \Ss(1,3)
          + 8 \* \Ss(1,4)
          - {29 \over 6} \* \Sss(1,1,2)
          - {2 \over 3} \* \S(2) \* \z3
  \nonumber\\&&
          - {20 \over 3} \* \Sss(1,3,1)
          + 12 \* \Ss(2,-3)
          - {236 \over 9} \* \Ss(2,-2)
          - {40 \over 3} \* \Sss(2,-2,1)
          - {4663 \over 1080} \* \Ss(2,1)
          + {32 \over 3} \* \Sss(2,1,-2)
          - {37 \over 3} \* \Sss(2,1,1)
          - {22817 \over 810} \* \S(3)
  \nonumber\\&&
          + {32 \over 3} \* \Ss(2,2)
          + {5 \over 3} \* \Ss(2,3)
          - 4 \* \Ss(3,-2)
          + {65 \over 18} \* \Ss(3,1)
          + {85 \over 6} \* \S(1) \* \z3
          \bigg]
     + (\Nminus+\Nplus)  \* \bigg[
            {86 \over 9} \* \Ss(1,-4)
          - {91 \over 36} \* \S(5)
          - 4 \* \Sss(3,1,1)
  \nonumber\\&&
          - {8 \over 9} \* \Ssss(1,-2,1,1)
          + {13226 \over 405} \* \Ss(1,-2)
          - {104 \over 9} \* \Sss(1,1,-3)
          + {152 \over 9} \* \Sss(1,1,-2)
          - {784 \over 27} \* \Ss(1,-3)
          - 4 \* \Sss(1,-2,-2)
          + {41929 \over 129600} \* \S(1)
  \nonumber\\&&
          - {244 \over 9} \* \S(1) \* \z3
          + {1 \over 2} \* \S(1) \* \z4
          + {208 \over 27} \* \Sss(1,-2,1)
          - {4 \over 9} \* \Sss(1,-3,1)
          - {13 \over 3} \* \Ss(1,1) \* \z3
          - {7 \over 8} \* \Sss(1,1,1)
          + {64 \over 9} \* \Ssss(1,1,1,-2)
          - {35 \over 6} \* \Ssss(1,1,1,1)
  \nonumber\\&&
          - {10 \over 3} \* \Sssss(1,1,1,1,1)
          + {44 \over 9} \* \Ssss(1,1,1,2)
          + {25 \over 3} \* \Sss(1,1,2)
          + 3 \* \Ssss(1,1,2,1)
          - {91 \over 18} \* \Sss(1,1,3)
          + {577 \over 648} \* \Ss(1,2)
          + {40 \over 3} \* \S(2) \* \z3
          + {107 \over 18} \* \Sss(1,2,1)
  \nonumber\\&&
          + {34 \over 9} \* \Ssss(1,2,1,1)
          - {79151 \over 64800} \* \S(2)
          - {68 \over 9} \* \Sss(1,2,-2)
          - {239 \over 54} \* \Ss(2,2)
          - {49 \over 9} \* \Sss(1,2,2)
          - {275 \over 108} \* \Ss(1,3)
          + {73 \over 18} \* \Ssss(2,1,1,1)
          - {31 \over 6} \* \Sss(2,1,2)
  \nonumber\\&&
          + {158 \over 9} \* \Ss(2,-3)
          - {754 \over 27} \* \Ss(2,-2)
          - {40 \over 9} \* \Sss(2,-2,1)
          + {133 \over 36} \* \Ss(4,1)
          - {403 \over 810} \* \Ss(2,1)
          - {68 \over 9} \* \Sss(2,1,-2)
          + {79 \over 18} \* \Sss(2,1,1)
          - {67 \over 18} \* \Sss(2,2,1)
  \nonumber\\&&
          + {17 \over 18} \* \Ss(1,4)
          + {5 \over 72} \* \S(4)
          - {5 \over 18} \* \Sss(1,3,1)
          + {25 \over 9} \* \Ss(2,3)
          + {7871 \over 3240} \* \S(3)
          + {134 \over 9} \* \Ss(3,-2)
          - {241 \over 72} \* \Ss(3,1)
          + {38 \over 9} \* \Ss(3,2)
          \bigg]
          \bigg)
\:\: .\label{eq:c2qq3}
\eea
For the sake of completeness, we include the result for the complete 
first and second-order longitudinal coefficient functions 
$c^{(1)}_{L,\rm ns}$ and $c^{(2)}_{L,\rm ns}$ known from 
refs.~\cite{Bardeen:1978yd,cLSanch,vanNeerven:1991nn,Moch:1999eb}
\bea
  &&c^{(1)}_{L,\rm ns}(N) \: = \:  
       - 4\, \* \cf \* (1-\Nplus)  \* \S(1)
\:\: ,\label{eq:cLqq1}
\\
  &&c^{(2)}_{L,\rm ns}(N) \: = \: 
 4\: \* \ca \* \cf \* \bigg(
            {12 \over 5} \* \z3 \* \delta(N\minus 2)
      + {12 \over 5} \* (\Nplus-\Nplustwo) \*  [ 
            \S(1)
          - \S(2)
          ]
      + {12 \over 5} \* (\Nplustwo-\Nplusthree) \*  [ 
            \Ss(1,-2)
          + \S(3)
          ]
  \quad \nonumber\\&&
          - {98 \over 15} \* \S(1)
          + {8 \over 5} \* \S(2)
          + {8 \over 5} \*(\Nminusthree-\Nminustwo) \* \Ss(1,-2)
          + 8 \* (\Nminus -1)  \* \Ss(1,-2)
      + {8 \over 5} \* (\Nminustwo-\Nminus)  \* [
            \S(1)
          + \S(2)
          ]
  \nonumber\\&&
      + (1-\Nplus)  \*  \bigg[
            12 \* \S(1) \* \z3
          - 4 \* \Ss(1,-2)
          - {23 \over 3} \* \Ss(1,1)
          - 8 \* \Sss(1,1,-2)
          - 4 \* \Ss(1,3)
          - {287 \over 18} \* \S(1)
          + 4 \* \Ss(1,-3)
          + {176 \over 15} \* \S(2)
          - 4 \* \S(3)
          \bigg]
  \nonumber\\&&
       + (\Nminus+\Nplus)  \*  \bigg[
            {49 \over 15} \* \S(1)
          - {4 \over 5} \* \S(2)
           \bigg]
          \bigg)
+ 4\: \* \cf \* \nf \* \bigg(
       (1-\Nplus)  \*  \bigg[
            {2 \over 3} \* \Ss(1,1)
          - {4 \over 3} \* \S(2)
          + {25 \over 9} \* \S(1)
          \bigg]
          - {2 \over 3} \* (\Nminus -1)  \* \S(1)
          \bigg)
  \nonumber\\&&
+ 4\: \* \cf^2 \*  \bigg(
          - {24 \over 5} \* \z3 \* \delta(N\minus 2) 
          + {22 \over 5} \* \S(1)
          + {4 \over 5} \* \S(2)
          - {16 \over 5} \*(\Nminusthree-\Nminustwo)  \* \Ss(1,-2)
     - {16 \over 5} \* (\Nminustwo-\Nminus)  \*  [
            \S(1)
          + \S(2)
          ]
  \nonumber\\&&
     + (\Nminus -1)  \*  [
            2 \* \Ss(1,1)
          - 16 \* \Ss(1,-2)
          ]
          - {24 \over 5} \* (\Nplus-\Nplustwo) \*  [ 
            \S(1)
          - \S(2)
          ]
     - {24 \over 5} \* (\Nplustwo-\Nplusthree)  \*  [
            \Ss(1,-2)
          + \S(3)
          ]
  \nonumber\\&&
     + (1-\Nplus)  \*  \bigg[
            {33 \over 2} \* \S(1)
          + 8 \* \Ss(1,3)
          - 24 \* \S(1) \* \z3
          - 8 \* \Ss(1,-3)
          + 8 \* \Ss(1,-2)
          + 7 \* \Ss(1,1)
          + 16 \* \Sss(1,1,-2)
          - 4 \* \Sss(1,1,1)
          + 4 \* \Ss(1,2)
  \nonumber\\&&
          - {54 \over 5} \* \S(2)
          + 6 \* \Ss(2,1)
          + 4 \* \S(3)
          \bigg]
       - (\Nminus+\Nplus)  \*  \bigg[
            {11 \over 5} \* \S(1)
          + {2 \over 5} \* \S(2)
          \bigg]
          \bigg)
\:\: .\label{eq:cLqq2}
\eea
The contributions to the 3-loop longitudinal coefficient 
function corresponding to eq.~(\ref{eq:c2qq3}) read
\bea
  &&c^{(3)}_{L,\rm ns}(N) \: = \: 
 16\: \* \ca \* \cf \* \nf \* \bigg(
         \delta(N\minus 2) \* \bigg[ 
            {20 \over 3} \* \z5
          - {149 \over 75} \* \z3
          \bigg]
+ (\Nminusthree - \Nminustwo)  \*  \bigg[
            {16 \over 15} \* \Ss(2,-2)
          - {4 \over 3} \* \Sss(1,3,1)
          + {4 \over 3} \* \Sss(1,1,3)
 \:\: \nonumber\\&&
          - {8 \over 3} \* \Ss(1,1) \* \z3
          - {8 \over 15} \* \Sss(1,1,-2)
          - {8 \over 15} \* \Sss(1,-2,1)
          + {8 \over 5} \* \Ss(1,-3)
          - {298 \over 225} \* \Ss(1,-2)
          +  {8 \over 5} \* \S(1) \* \z3
          \bigg]
+ (\Nminustwo - \Nminus)  \*  \bigg[
            {4 \over 3} \* \Ss(1,1) \* \z3
  \nonumber\\&&
          - {418 \over 225} \* \S(1)
          + {8 \over 5} \* \S(3)
          + {8 \over 15} \* \Ss(1,-2)
          + {4 \over 3} \* \Sss(1,-2,1)
          - {4 \over 3} \* \Sss(1,1,-2)
          + {4 \over 5} \* \Ss(1,1)
          - {178 \over 225} \* \S(2)
          + {4 \over 3} \* \Ss(3,1)
          + 4 \* \S(1) \* \z3
          + {4 \over 5} \* \Ss(2,1)
  \nonumber\\&&
          - {2 \over 3} \* \Sss(1,1,3)
          - {4 \over 3} \* \Ss(1,3)
          + {2 \over 3} \* \Sss(1,3,1)
          \bigg]
+ (\Nminus -1) \*  \bigg[
            {16 \over 3} \* \Ss(2,-2)
          - {13033 \over 1350} \* \S(1)
          + 8 \* \S(1) \* \z3
          - {518 \over 45} \* \Ss(1,-2)
          - {2 \over 5} \* \Ss(2,1)
  \nonumber\\&&
          - {8 \over 3} \* \Sss(1,-2,1)
          - {254 \over 45} \* \Ss(1,1)
          - {2 \over 15} \* \S(3)
          +  8 \* \Ss(1,-3)
          + {76 \over 25} \* \S(2)
          - {8 \over 3} \* \Sss(1,1,-2)
          \bigg]
+ (\Nplus-\Nplustwo)  \*  \bigg[
            {8 \over 3} \* \Sss(1,-2,1)
          - {8 \over 3} \* \S(2) \* \z3
  \nonumber\\&&
          - {7 \over 25} \* \S(2)
          - {8 \over 3} \* \Sss(1,1,-2)
          - 2 \* \Ss(1,3)
          + {4 \over 3} \* \Sss(1,3,1)
          - {4 \over 3} \* \Sss(1,1,3)
          + 2 \* \Ss(1,2)
          + {16 \over 5} \* \Ss(1,1)
          + {8 \over 3} \* \Ss(1,1) \* \z3
          + {4 \over 5} \* \Ss(1,-2)
          - {16 \over 5} \* \Ss(2,1)
  \nonumber\\&&
          + {20 \over 3} \* \S(1) \* \z3
          - {73 \over 25} \* \S(1)
          - {2 \over 5} \* \S(3)
          - {2 \over 3} \* \Ss(3,1)
          - {4 \over 3} \* \Ss(4,1)
          + {4 \over 3} \* \Ss(2,3)
          + {8 \over 3} \* \Ss(2,-2)
          \bigg]
+ (\Nplustwo-\Nplusthree)  \*  \bigg[
            {4 \over 5} \* \Ss(2,-2)
          + 2 \* \Ss(2,2)
  \nonumber\\&&
          - {4 \over 5} \* \Sss(1,-2,1)
          - 2 \* \Ss(2,3)
          - {14 \over 5} \* \Ss(3,1)
          - {53 \over 25} \* \S(3)
          + {12 \over 5} \* \S(4)
          + 2 \* \Ss(4,1)
          - {48 \over 5} \* \S(1) \* \z3
          + {12 \over 5} \* \Ss(1,-3)
          + 2 \* \Sss(1,1,3)
          - 4 \* \Ss(1,1) \* \z3
  \nonumber\\&&
          - {4 \over 5} \* \Sss(1,1,-2)
          - {53 \over 25} \* \Ss(1,-2)
          + 2 \* \Sss(1,2,1)
          - 2 \* \Sss(1,3,1)
          + 4 \* \S(2) \* \z3
          - 2 \* \Sss(1,1,2)
          \bigg]
+ (1-\Nplus)  \*  \bigg[
            {125599 \over 4050} \* \S(1)
          - {40 \over 3} \* \S(1) \* \z3
  \nonumber\\&&
          + 4 \* \Ss(1,-4)
          - {68 \over 9} \* \Ss(1,-3)
          + {413 \over 45} \* \Ss(1,-2)
          - {8 \over 3} \* \Sss(1,-2,-2)
          + {4469 \over 270} \* \Ss(1,1)
          - 8 \* \Ss(1,1) \* \z3
          - 8 \* \Sss(1,1,-3)
          + {88 \over 9} \* \Sss(1,1,-2)
  \nonumber\\&&
          + {40 \over 9} \* \Sss(1,1,1)
          + {16 \over 3} \* \Ssss(1,1,1,-2)
          + {2 \over 3} \* \Ssss(1,1,1,2)
          + {2 \over 3} \* \Sss(1,1,2)
          - {2 \over 3} \* \Ssss(1,1,2,1)
          + {2 \over 3} \* \Sss(1,1,3)
          - {40 \over 9} \* \Ss(1,2)
          - {16 \over 3} \* \Sss(1,2,-2)
  \nonumber\\&&
          - {10 \over 3} \* \S(4)
          - {2 \over 3} \* \Sss(1,2,1)
          - {2 \over 3} \* \Sss(1,2,2)
          + {2 \over 3} \* \Ss(3,1)
          + {8 \over 3} \* \Sss(1,3,1)
          - 4 \* \Ss(1,4)
          - {20953 \over 675} \* \S(2)
          + 8 \* \S(2) \* \z3
          + {4 \over 3} \* \Ss(2,-3)
          - {2 \over 3} \* \Ss(2,-2)
  \nonumber\\&&
          + {8 \over 3} \* \Sss(2,-2,1)
          - {238 \over 45} \* \Ss(2,1)
          - {16 \over 3} \* \Sss(2,1,-2)
          - 2 \* \Ss(2,3)
          + {824 \over 45} \* \S(3)
          + {38 \over 9} \* \Ss(1,3)
          + {2 \over 3} \* \Ss(4,1)
          \bigg]
          \bigg)
  \nonumber\\&&
+ 16\: \* \cf \* \nf^2 \* \bigg(
 (\Nminus-1)  \* \bigg[ 
            {19 \over 27} \* \S(1)
          + {2 \over 9} \* \Ss(1,1)
          - {2 \over 9} \* \S(2)
          \bigg]
+ (1-\Nplus)  \* \bigg[ 
            {2 \over 9} \* \Ss(1,2)
          - {317 \over 162} \* \S(1)
          - {25 \over 27} \* \Ss(1,1)
          - {2 \over 9} \* \Sss(1,1,1)
  \nonumber\\&&
          + {50 \over 27} \* \S(2)
          + {4 \over 9} \* \Ss(2,1)
          - {2 \over 3} \* \S(3)
          \bigg]
          \bigg)
+ 16\: \* \cf^2 \* \nf \*  \bigg(
         {298 \over 75} \* \z3 \* \delta(N\minus 2) 
 + (\Nminusthree - \Nminustwo)   \*   \bigg[
            {16 \over 15} \* \Sss(1,1,-2)
          - {16 \over 5} \* \Ss(1,-3)
  \nonumber\\&&
          + {16 \over 15} \* \Sss(1,-2,1)
          + {596 \over 225} \* \Ss(1,-2)
          - {32 \over 15} \* \Ss(2,-2)
          - {16 \over 5} \* \S(1) \* \z3
          \bigg]
 + (\Nminustwo - \Nminus)  \*   \bigg[
            {16 \over 15} \* \Ss(2,1)
          - {16 \over 5} \* \S(3)
          + {356 \over 225} \* \S(2)
  \nonumber\\&&
          + {836 \over 225} \* \S(1)
          - {8 \over 3} \* \S(1) \* \z3
          + {16 \over 15} \* \Ss(1,1)
          - {8 \over 3} \* \Sss(1,-2,1)
          + {8 \over 3} \* \Sss(1,1,-2)
          - {16 \over 15} \* \Ss(1,-2)
          \bigg]
 + (\Nminus - 1)  \*   \bigg[
            {2141 \over 1350} \* \S(1)
          - 16 \* \S(1) \* \z3
  \nonumber\\&&
          - 16 \* \Ss(1,-3)
          + {1036 \over 45} \* \Ss(1,-2)
          + {16 \over 3} \* \Sss(1,-2,1)
          + {73 \over 45} \* \Ss(1,1)
          + {16 \over 3} \* \Sss(1,1,-2)
          - {5 \over 3} \* \Sss(1,1,1)
          + {5 \over 3} \* \Ss(1,2)
          + {232 \over 225} \* \S(2)
          + {1 \over 10} \* \S(3)
  \nonumber\\&&
          - {32 \over 3} \* \Ss(2,-2)
          + {14 \over 5} \* \Ss(2,1)
          \bigg]
 + (\Nplus-\Nplustwo)  \*   \bigg[
            {146 \over 25} \* \S(1)
          - {10 \over 3} \* \Ss(1,2)
          - {308 \over 75} \* \S(2)
          + {20 \over 3} \* \Ss(3,1)
          - {4 \over 3} \* \Ss(2,2)
          + {8 \over 3} \* \S(1) \* \z3
  \nonumber\\&&
          - {16 \over 3} \* \Ss(2,-2)
          - {4 \over 3} \* \Sss(1,2,1)
          + {16 \over 3} \* \Sss(1,1,-2)
          + {4 \over 3} \* \Sss(1,1,2)
          - {8 \over 5} \* \Ss(1,-2)
          - {16 \over 3} \* \Sss(1,-2,1)
          - {26 \over 15} \* \Ss(1,1)
          + {24 \over 5} \* \S(3)
          + {26 \over 15} \* \Ss(2,1)
          \bigg]
  \nonumber\\&&
 + (\Nplustwo-\Nplusthree)  \*   \bigg[
            {76 \over 5} \* \S(1) \* \z3
          + {106 \over 25} \* \Ss(1,-2)
          + {8 \over 5} \* \Sss(1,-2,1)
          - {24 \over 5} \* \Ss(1,-3)
          + {10 \over 3} \* \Sss(1,1,2)
          + {8 \over 5} \* \Sss(1,1,-2)
          - {10 \over 3} \* \Sss(1,2,1)
  \nonumber\\&&
          - {8 \over 5} \* \Ss(2,-2)
          + {74 \over 15} \* \Ss(3,1)
          - {24 \over 5} \* \S(4)
          + {106 \over 25} \* \S(3)
          - {10 \over 3} \* \Ss(2,2)
          \bigg]
 + (1-\Nplus)  \*   \bigg[
            {74 \over 3} \* \S(1) \* \z3
          - {173653 \over 10800} \* \S(1)
          - 8 \* \Ss(1,-4)
  \nonumber\\&&
          + {136 \over 9} \* \Ss(1,-3)
          - {826 \over 45} \* \Ss(1,-2)
          + {16 \over 3} \* \Sss(1,-2,-2)
          - {1609 \over 135} \* \Ss(1,1)
          + {40 \over 3} \* \Ss(1,1) \* \z3
          + 16 \* \Sss(1,1,-3)
          - {176 \over 9} \* \Sss(1,1,-2)
  \nonumber\\&&
          + {7 \over 18} \* \Sss(1,1,1)
          - {32 \over 3} \* \Ssss(1,1,1,-2)
          + {8 \over 3} \* \Ssss(1,1,1,1)
          - {4 \over 3} \* \Ssss(1,1,1,2)
          + 8 \* \Ss(1,4)
          - 4 \* \Sss(1,1,2)
          + {4 \over 3} \* \Ssss(1,1,2,1)
          - {23 \over 18} \* \Ss(1,2)
          + {17 \over 6} \* \S(4)
  \nonumber\\&&
          + {32 \over 3} \* \Sss(1,2,-2)
          - 2 \* \Sss(1,2,1)
          + {4 \over 3} \* \Sss(1,2,2)
          - {49 \over 9} \* \Ss(1,3)
          - {20 \over 3} \* \Sss(1,3,1)
          + {30737 \over 1350} \* \S(2)
          - {40 \over 3} \* \S(2) \* \z3
          - {8 \over 3} \* \Ss(2,-3)
          + {4 \over 3} \* \Ss(2,-2)
  \nonumber\\&&
          - {16 \over 3} \* \Sss(2,-2,1)
          - {17 \over 10} \* \Ss(2,1)
          + {32 \over 3} \* \Sss(2,1,-2)
          - {13 \over 3} \* \Sss(2,1,1)
          + {13 \over 3} \* \Ss(2,2)
          + {8 \over 3} \* \Ss(2,3)
          - {913 \over 45} \* \S(3)
          + {13 \over 3} \* \Ss(3,1)
          \bigg]
          \bigg)
\:\: .\label{eq:cLqq3}
\eea
Recall that in all our formulae the expansion parameter is normalized 
as in eq.~(\ref{eq:Cexp}). The operators $\Npm$ and $\Npmi$ have been 
defined in eq.~(\ref{eq:shiftN}).
\end{appendix}
%
%
\end{document}